\documentclass[preprint,aps]{revtex4}
\usepackage{graphicx}
\usepackage{amssymb}
\newcommand{\be}{\begin{eqnarray}}
\newcommand{\ee}{\end{eqnarray}}
\usepackage{hyperref}
\usepackage{cancel} 
\newcommand{\nn}{\nonumber}

\usepackage{amsmath,amssymb}

\begin{document}

\title{Quark Wigner Distributions and Orbital Angular Momentum in Light-front
Dressed Quark Model}

\author{\bf Asmita Mukherjee, Sreeraj Nair and Vikash Kumar Ojha}

\affiliation{ Department of Physics,
Indian Institute of Technology Bombay,\\ Powai, Mumbai 400076,
India.}
\date{\today}

\begin{abstract}
\noindent
We calculate the Wigner functions for a quark target dressed with a gluon at
one loop in perturbation theory. The Wigner distributions give a combined position 
and momentum space information of the quark
distributions and are related to both generalized parton distributions
(GPDs)  and transverse momentum dependent parton distributions (TMDs). 
We calculate and compare the different definitions of quark orbital angular
momentum and the spin-orbit correlations in this perturbative model. We compare our 
results with other model calculations.  
\end{abstract}

\maketitle

\section{Introduction}
\noindent
In classical physics, a system of particles can be described in terms of
phase space distributions, which represent the density of particles at a
point in the phase space at a given time. In quantum mechanics, position and
momentum operators do not commute and they cannot be determined
simultaneously. Thus in quantum mechanics one cannot define phase space
distributions. Wigner distributions in quantum mechanics have been
introduced long ago \cite{wigner}, which can be thought of as quantum mechanical phase
space distributions, however it cannot be interpreted as probability
distribution for the reason above and is not positive definite. Wigner
distributions become classical phase space distributions in the limit
$h \rightarrow 0$. A quantum mechanical Wigner distribution for the quarks and gluons
in the rest frame of the nucleon was introduced in \cite{ji1,ji2}. Reduced Wigner
functions are obtained from the seven dimensional most general Wigner
distributions by integrating the minus component of the momentum. Reduced
Wigner distributions are functions of three position and three momentum
variables and as discussed above are not measurable. To obtain measurable
quantities one has to integrate over more variables. Integrating out the
momentum variables one can relate the reduced Wigner distributions to
generalized parton distributions (GPDs) and integrating out the position
variables one gets the transverse momentum dependent parton distributions
(TMDs). Thus the Wigner distributions can be thought of as more general
mother distributions in which both position and momentum space information
of quarks and gluons are encoded.

Wigner distributions are related to the generalized transverse momentum
dependent correlation functions (GPCFs) \cite{metz,spinhalf} of the nucleon, 
which are the fully
unintegrated, off-diagonal quark-quark correlators. An overlap
representation for the above using model light-front wave functions has been
studied in \cite{unified}. If one integrates over
the minus component of the momentum (light-cone energy) one gets the
generalized transverse momentum dependent parton distributions (GTMDs).
These are functions of the 3-momentum of the quark and the momentum transfer 
to the nucleon $\Delta_\mu$. In \cite{lorce} the authors introduced five
dimensional Wigner distributions in infinite momentum frame by integrating
the GTMDs over the momentum transfer in the transverse direction
$\Delta_\perp$. These Wigner distributions are functions of the two position
and three momentum variables. Working in infinite momentum frame, or
equivalently using the light-cone formalism has several advantages as the
transverse boosts are Galilean, or do not involve dynamics, and longitudinal
boost is just a scale transformation \cite{pedestrian}. So it is easier 
to have an intuitive picture of the parton distributions in the nucleon. 
As discussed before, Wigner distributions do not have probabilistic
interpretation due to uncertainty principle. Integrating out one or more
variables one can define new distributions that have probabilistic
interpretation. Depending on whether the nucleon and the quark is polarized
or unpolarized several such distributions can be defined. In this work we
shall restrict ourselves to longitudinal polarization's only. As Wigner
distributions cannot be measured, model calculations are of importance to
understand what kind of information about the quark-gluon correlation in the
nucleon can be obtained from them as well as to verify to what extent
different model dependent and model independent relations among various
distributions are satisfied. However, integrating out more variables 
gives measurable quantities having the interpretation of probability
densities. In \cite{lorce} the Wigner distributions for quarks and gluons have been
studied in light cone constituent quark model and in light-cone chiral quark
soliton model. Both these models have no gluonic degrees of freedom and the
Wilson line becomes unity. 

Quark orbital angular momentum (OAM) contribution to the total spin of the nucleon 
has gained considerable attention since the EMC experiments \cite{emc} which showed that 
the quark intrinsic spin contribution was less than expected. Also recent 
polarized beam experiments suggest that the gluon polarization contribution to the total
spin of the proton is very small.
Wigner distributions are related to the OAM
carried by the quarks in the nucleon. As suggested from the experimental
data, a substantial part
of the spin of the nucleon comes from quark and gluon OAM. The issue of
gauge invariance and experimental measurability of the OAM contribution 
complicates the issue of a full understanding of such contributions
\cite{OAM_rev}.
Theoretically there exist mainly two definitions of OAM : one obtained from 
the sum rules of GPDs and the other, canonical OAM distribution in the
light cone gauge. It has been shown in the literature that these two
different distributions are projections of Wigner distributions with
different choice of gauge links and they are related by a gauge dependent
potential term \cite{mb,hatta1,wilson}. In \cite{hatta,ji_or} the canonical OAM 
in light-front gauge is shown to
be related to the twist three GPDs.

In this paper, we present a calculation of the quark Wigner distributions
in light-front Hamiltonian formulation using overlaps of light-front wave
functions (LFWFs). This approach is based on \cite{zhang}. This has the
advantage that it gives an intuitive picture of deep inelastic scattering
(DIS) processes in field theory while keeping close contact with parton
model, but the partons are now field theoretic partons, they are
non-collinear, massive and also interacting \cite{kundu1}. However, they are still on-mass
shell. An expansion of the target state in Fock space in terms of
multi-parton LFWFs allows one to calculate the matrix
elements of operators. The non-perturbative light-front wave functions are
boost invariant. While the non-perturbative LFWFs for a bound state like the
nucleon requires a model light-front Hamiltonian, it is interesting and
useful to replace the bound state by a simple composite two-body spin
$1/2$ state, like a quark at one loop in perturbation theory. This is a
relativistic state and the relativistic two-parton LFWFs can be calculated
analytically in light-front Hamiltonian perturbation theory. These wave
function is  a function of the mass of the quark. It mimics the 
LFWF of a two-particle bound state \cite{brodsky_OAM}. In this work we
calculate the Wigner distributions and OAM for a quark dressed with a gluon
in light-front Hamiltonian approach. We follow the
formalism of \cite{zhang} where it was shown that in light-front gauge one can 
write the light-front QCD Hamiltonian entirely in terms of the dynamical
degrees of freedom and using a certain representation
of the Dirac gamma matrices, it is possible to  write the theory in terms
of two-component fermion spinors and transverse components of the gauge
field. This two-component approach has been used successfully to investigate
the GPDs. Here we use this formalism to investigate the Wigner
distributions.               

The plan of the paper is as follows. In section II we calculate the Wigner
distributions for a dressed quark. In section III we calculate the OAM
in the same model. We present the numerical results in section IV and
conclusions in section V. 
\section{Wigner Distributions}
\noindent
The Wigner distribution of quarks can be defined as the two-dimensional Fourier transforms 
of the generalized transverse momentum distributions (GTMDs) 
\cite{lorce,metz} 

\be \label{main}
\rho^{[\Gamma]} ({b}_{\perp},{k}_{\perp},x,\sigma) = \int \frac{d^2 \Delta_{\perp}}
{(2\pi)^2} e^{-i \Delta_{\perp}.b_{\perp}}
W^{[\Gamma]} (\Delta_{\perp},{k}_{\perp},x,\sigma);
\ee 
 where  $\Delta_{\perp}$ is momentum transfer of dressed quark in transverse direction and 
${b}_{\perp}$ is 2 dimensional vector
 in impact parameter space conjugate to $\Delta_{\perp}$. $W^{[\Gamma]}$ is the quark-quark 
correlator given by
\be \label{qqc}
W^{[\Gamma]} ({\Delta}_{\perp},{k}_{\perp},x,\sigma)  =  \Big{\langle }p^{+},
\frac{\Delta_{\perp}}{2},\sigma  \Big{|}
W^{[\Gamma]} (0_{\perp},k_{\perp},x)  \Big{|}  p^{+},-\frac{\Delta_{\perp}}{2},\sigma \Big{\rangle } 
 \nn \\ \nn \\
=\frac{1}{2}\int\frac{dz^{-}d^{2} z_{\perp}}{(2\pi)^3}e^{i(xp^+ z^-/2-k_{\perp}.z_{\perp})}
 \Big{\langle } p^{+},\frac{\Delta_{\perp}}{2},\sigma \Big{|}
\overline{\psi}(-\frac{z}{2}) \Omega\Gamma \psi(\frac{z}{2}) \Big{|}
p^{+},-\frac{\Delta_{\perp}}{2},\sigma \Big{\rangle }
\Big{|}_{z^{+}=0}.
\ee \\
We define the initial and final dressed quark state in the symmetric frame \cite{brodsky} 
where $p^+$ and 
$\sigma$ defines the longitudinal momentum of the target state and its
helicity respectively. $x = k^+/p^+$ is the fraction of 
longitudinal momentum of the dressed quark carried by the quark. In the symmetric frame the transverse momentum transfer($\Delta_{\perp}$) has the 
$\Delta_{\perp} \longrightarrow -\Delta_{\perp}$ symmetry. $\Omega$ is the
gauge link needed for color gauge invariance. In this work, we use the
light-front gauge and take the gauge link to be unity. The symbol $\Gamma$ represents 
the Dirac matrix defining the
types of quark densities.\\
\\%
In this work, we calculate the above Wigner distributions for  a quark state
dressed with a gluon. The state of momentum $p$ and helicity $\sigma$, 
can be expanded in Fock space in terms of
multi-parton light-front wave functions (LFWFs) \cite{hari}   
\be \label{dqs}
  \Big{| }p^{+},p_{\perp},\sigma  \Big{\rangle} = \Phi^{\sigma}(p) b^{\dagger}_{\sigma}(p)
 | 0 \rangle +
 \sum_{\sigma_1 \sigma_2} \int [dp_1]
 \int [dp_2] \sqrt{16 \pi^3 p^{+}}
 \delta^3(p-p_1-p_2) \nn \\ \Phi^{\sigma}_{\sigma_1 \sigma_2}(p;p_1,p_2) 
b^{\dagger}_{\sigma_1}(p_1) 
 a^{\dagger}_{\sigma_2}(p_2)  | 0 \rangle;
\ee
where $[dp] =  \frac{dp^{+}d^{2}p_{\perp}}{ \sqrt{16 \pi^3 p^{+}}}$. $ \Phi^{\sigma}(p)$ and
$ \Phi^{\sigma}_{\sigma_1 \sigma_2}$ are the single particle (quark) and two particle (quark-gluon) light-front wave function (LFWF).
$\sigma_1$ and $\sigma_2$ are the helicities of the quark and gluon respectively. 
$ \Phi^{\sigma}(p)$ gives  the wave function renormalization for the quark. 
The two particle
function $ \Phi^{\sigma}_{\sigma_1 \sigma_2}(p;p_1,p_2)$ gives the probability to find a bare 
quark having momentum $p_1$ and helicity $\sigma_1$ and a bare gluon with momentum $p_2$ and 
helicity $\sigma_2$ in the dressed quark. The two particle LFWF 
is related to the boost invariant LFWF; $\Psi^{\sigma}_{\sigma_1
\sigma_2}(x, q_\perp) =   
\Phi^{\sigma}_{\sigma_1 \sigma_2}
\sqrt{P^+}$.  Here we have used the Jacobi momenta $(x_i, q_{i \perp})$ : 
\be 
p_i^+= x_i p^+, ~~~~~~~~~~q_{i \perp}= k_{i \perp}+x_i p_\perp
\ee
so that $\sum_i x_i=1, \sum_i q_{i\perp}=0$.   
These two-particle LFWFs  be calculated perturbatively as \cite{hari}:
\be \label{tpa}
\Psi^{\sigma a}_{\sigma_1 \sigma_2}(x,q_{\perp}) = 
\frac{1}{\Big[    m^2 - \frac{m^2 + (q_{\perp})^2 }{x} - \frac{(q_{\perp})^2}{1-x} \Big]}
\frac{g}{\sqrt{2(2\pi)^3}} T^a \chi^{\dagger}_{\sigma_1} \frac{1}{\sqrt{1-x}}
\nn \\ \Big[ 
-2\frac{q_{\perp}}{1-x}   -  \frac{(\sigma_{\perp}.q_{\perp})\sigma_{\perp}}{x}
+\frac{im\sigma_{\perp}(1-x)}{x}\Big]
\chi_\sigma (\epsilon_{\perp \sigma_2})^{*}.
\ee
We use the two component formalism \cite{zhang}; $\chi$ is the two
component spinor, $T^a$ are the color $SU(3)$ matrices, $m$ is 
the mass of the quark and $\epsilon_{\perp \sigma_2}$ is the polarization
vector of the gluon; $\perp=1,2$.  As stated in the introduction, the quark state dressed by
a gluon as we consider here mimics the bound state of a spin-1/2 particle
and a spin-1 particle. For such a bound state the bound state mass $M$
should be less than the sum of the masses of the constituents for stability. 
Here in the two-component formalism, we use the same mass for the bare
as well as the dressed quark in perturbation theory \cite{kundu1}. We investigate the 
Wigner distributions for
unpolarized and longitudinally polarized dressed quark and the relevant
correlators are with $\Gamma = \gamma^+$ and $\gamma^+
\gamma_5$. The single particle sector contributes through the normalization
of the state, which is important to get the complete contribution at $x=1$.
In this work we restrict ourselves to the kinematic region $x<1$, and in
this case the contribution from $ \Phi^{\sigma}(p)$ can be taken to be $1$. 
We calculate the contribution to the quark-quark correlator and
the Wigner distribution from the two particle sector in 
the Fock space expansion. This  is given by 
\be \label{w1}
W^{[  \gamma^{+}]} (\Delta_{\perp},k_{\perp},x,\sigma)  = 
\frac{1}{(2\pi)^3}
 \sum_{\sigma_1, \sigma_{2}} 
 \Psi^{*\sigma a}_{\sigma_{1} \sigma_2}(x,q'_{\perp}) 
\Psi^{\sigma a}_{\sigma_1 \sigma_2}(x,q_{\perp}),
\ee 

\be \label{w2}
W^{[  \gamma^{+}\gamma_5]} (\Delta_{\perp},k_{\perp},x,\sigma)  = 
\frac{1}{(2\pi)^3}
 \sum_{\sigma_1, \sigma_{2},\lambda_{1}} 
 \Psi^{*\sigma a}_{\lambda_{1} \sigma_2}(x,q'_{\perp}) 
\chi_{\lambda_1}^{\dagger}\sigma_3 \chi_{\sigma_1}
\Psi^{\sigma a}_{\sigma_1 \sigma_2}(x,q_{\perp});
\ee \\%
where the Jacobi relation for the transverse momenta in the symmetric frame is given by 
$q'_{\perp} = k_{\perp}-\frac{\Delta_{\perp}}{2}(1-x)$ and $q_{\perp} 
= k_{\perp}+\frac{\Delta_{\perp}}{2}(1-x)$.
We use the symbol $\rho_{\lambda\lambda^\prime}$ for Wigner distributions, where 
$\lambda(\lambda^\prime)$ is 
longitudinal polarization of target state(quark).
The four Wigner distributions have been defined in \cite{lorce} as

\be \label{ruu}
\rho_{UU}({b}_\perp,{k}_\perp,x) = \frac{1}{2}\Big[\rho^{[\gamma^+]}
({b}_\perp,{k}_\perp,x,+{e}_z) +
\rho^{[\gamma^+]}({b}_\perp,{k}_\perp,x,-{e}_z) \Big]
\ee
is the Wigner distribution of unpolarized quarks in unpolarized target state.

\be \label{rlu}
\rho_{LU}({b}_\perp,{k}_\perp,x) = \frac{1}{2}\Big[\rho^{[\gamma^+]}
({b}_\perp,{k}_\perp,x,+{e}_z) -
\rho^{[\gamma^+]}({b}_\perp,{k}_\perp,x,-{e}_z) \Big]
\ee
is the distortion due to longitudinal polarization of  the target state.

\be \label{rul}
\rho_{UL}({b}_\perp,{k}_\perp,x) = \frac{1}{2}\Big[\rho^{[\gamma^+\gamma_5]}
({b}_\perp,{k}_\perp,x,+{e}_z)+
\rho^{[\gamma^+\gamma_5]}({b}_\perp,{k}_\perp,x,-{e}_z) \Big]
\ee
represents distortion due to the longitudinal polarization of quarks, and

\be \label{rll}
\rho_{LL}({b}_\perp,{k}_\perp,x) = \frac{1}{2}\Big[\rho^{[\gamma^+\gamma_5]}
({b}_\perp,{k}_\perp,x,+{e}_z)-
\rho^{[\gamma^+\gamma_5]}({b}_\perp,{k}_\perp,x,-{e}_z) \Big]
\ee
represents the distortion due to the correlation between the longitudinal
polarized target state and quarks. \\%
In our case, $+{e_z}$ and $-{e_z}$ correspond to helicity up and
down of the target state, respectively.  In the model we consider, $\rho_{LU} = \rho_{UL}$
and the final expression for the three independent Wigner distribution 
are as follows:
\be\label{rhouu}
\rho^{[\gamma^{+}]}_{UU} (b_{\perp},k_{\perp},x) = 
N \int d \Delta_x \int d \Delta_y  
\frac{ \mathrm{ cos}(\Delta_\perp\cdot b_\perp)}{D(q_{\perp})D(q'_{\perp})}
\Big[I_1+\frac{4m^2(1-x)}{x^2}\Big];
\ee
\be\label{rholu}
\rho^{[\gamma^{+}]}_{LU} (b_{\perp},k_{\perp},x) & = &
N \int d \Delta_x \int d \Delta_y  
\frac{ \mathrm{ sin}(\Delta_\perp\cdot b_\perp)}{D(q_{\perp})D(q'_{\perp})} \Big[4( k_x \Delta_y - k_y \Delta_x ) 
\frac{(1+x)}{x^2(1-x)} \Big];
\ee
\be\label{rholl}
\rho^{[\gamma^{+}\gamma_{5}]}_{LL} (b_{\perp},k_{\perp},x) &= &
N \int d \Delta_x \int d \Delta_y  
\frac{ \mathrm{ cos}(\Delta_\perp\cdot b_\perp)}{D(q_{\perp})D(q'_{\perp})}
\Big[I_1-\frac{4m^2(1-x)}{x^2}\Big]; 
\ee

where $A_x ,A_y$ are $x,y$ component of $A_\perp$ and
\be
D(k_{\perp})=   \Big(m^2 - \frac{m^2 + (k_{\perp})^2 }{x} - \frac{(k_{\perp})^2}{1-x}\Big)\hspace{1cm}
I_1 = 4 \Big((k_{\perp})^2 - \frac{ \Delta_{\perp}^{2}(1-x)^2}{4} \Big)
 \frac{(1+x^2)}{x^2(1-x)^3}. 
 \nn \ee
Wigner distribution are real \cite{lorce}, which is due to the
Hermiticity property of the GTMDs to which they are related; and in the
above expressions, we have
taken the real part of the Fourier transforms.

\begin{figure}[!htp]
\begin{minipage}[c]{1\textwidth}
\tiny{(a)}\includegraphics[width=7.8cm,height=5cm,clip]{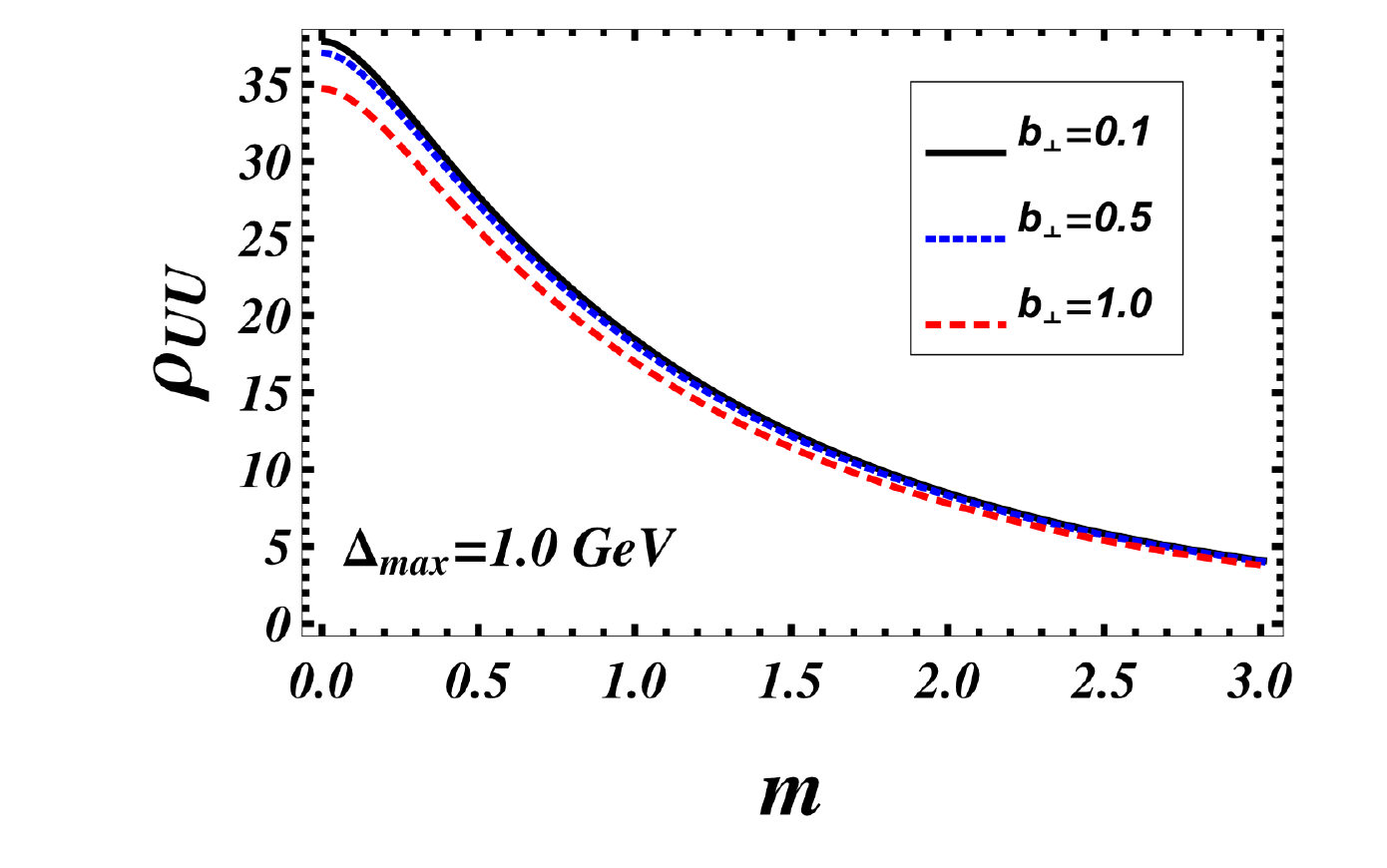}
\hspace{0.1cm}
\tiny{(b)}\includegraphics[width=7.8cm,height=5cm,clip]{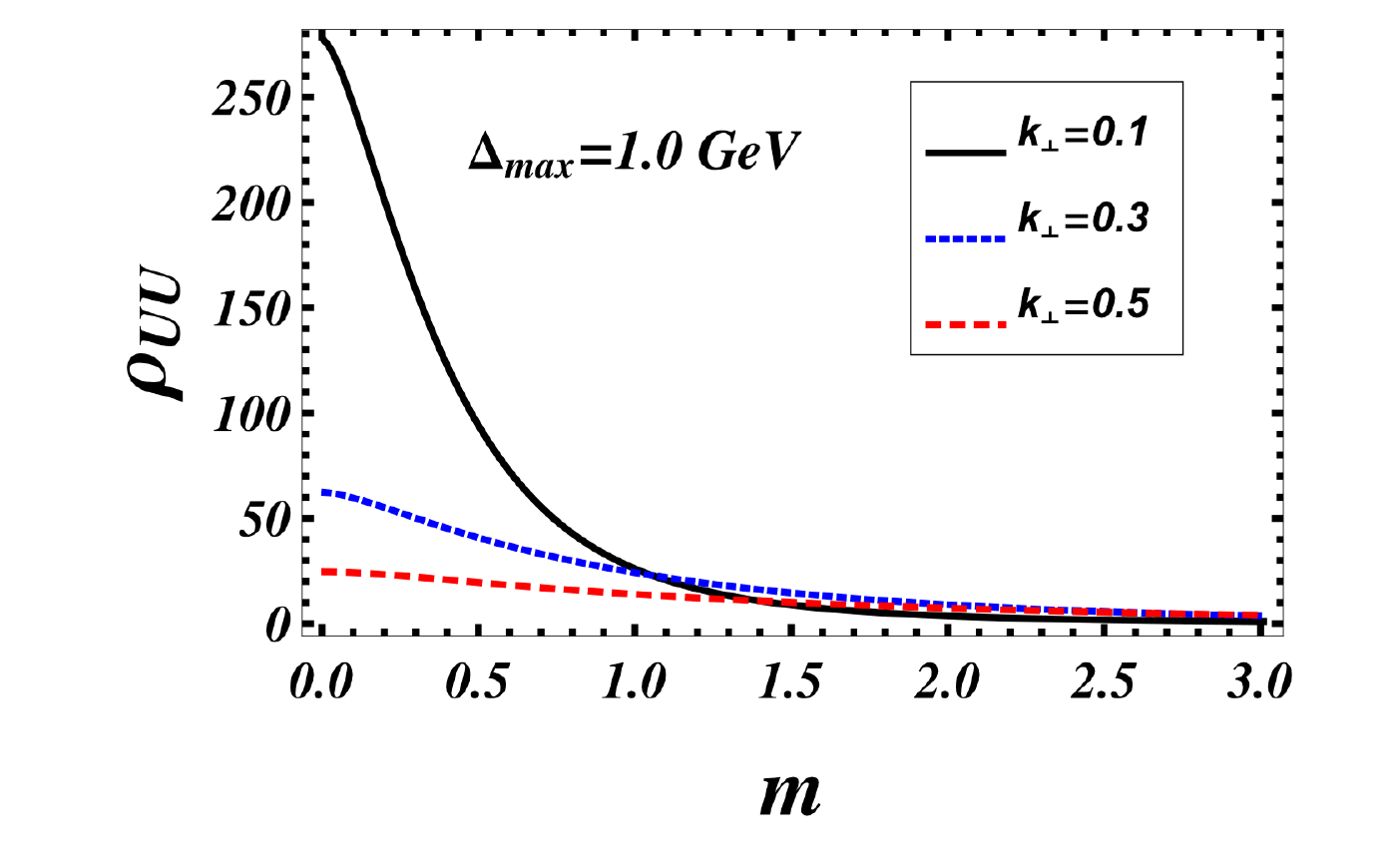}
\end{minipage}
\begin{minipage}[c]{1\textwidth}
\tiny{(c)}\includegraphics[width=7.8cm,height=5cm,clip]{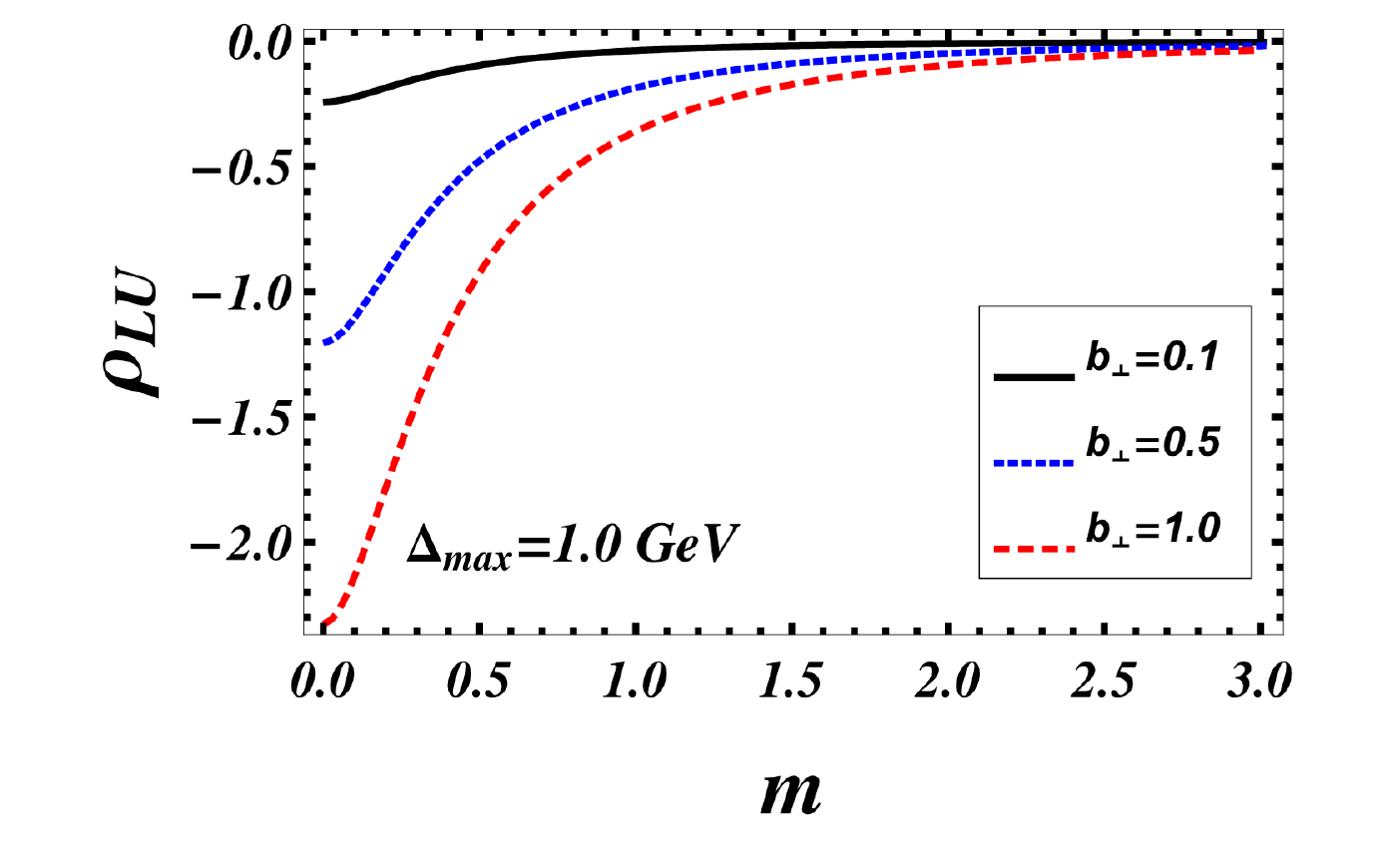}
\hspace{0.1cm}
\tiny{(d)}\includegraphics[width=7.8cm,height=5cm,clip]{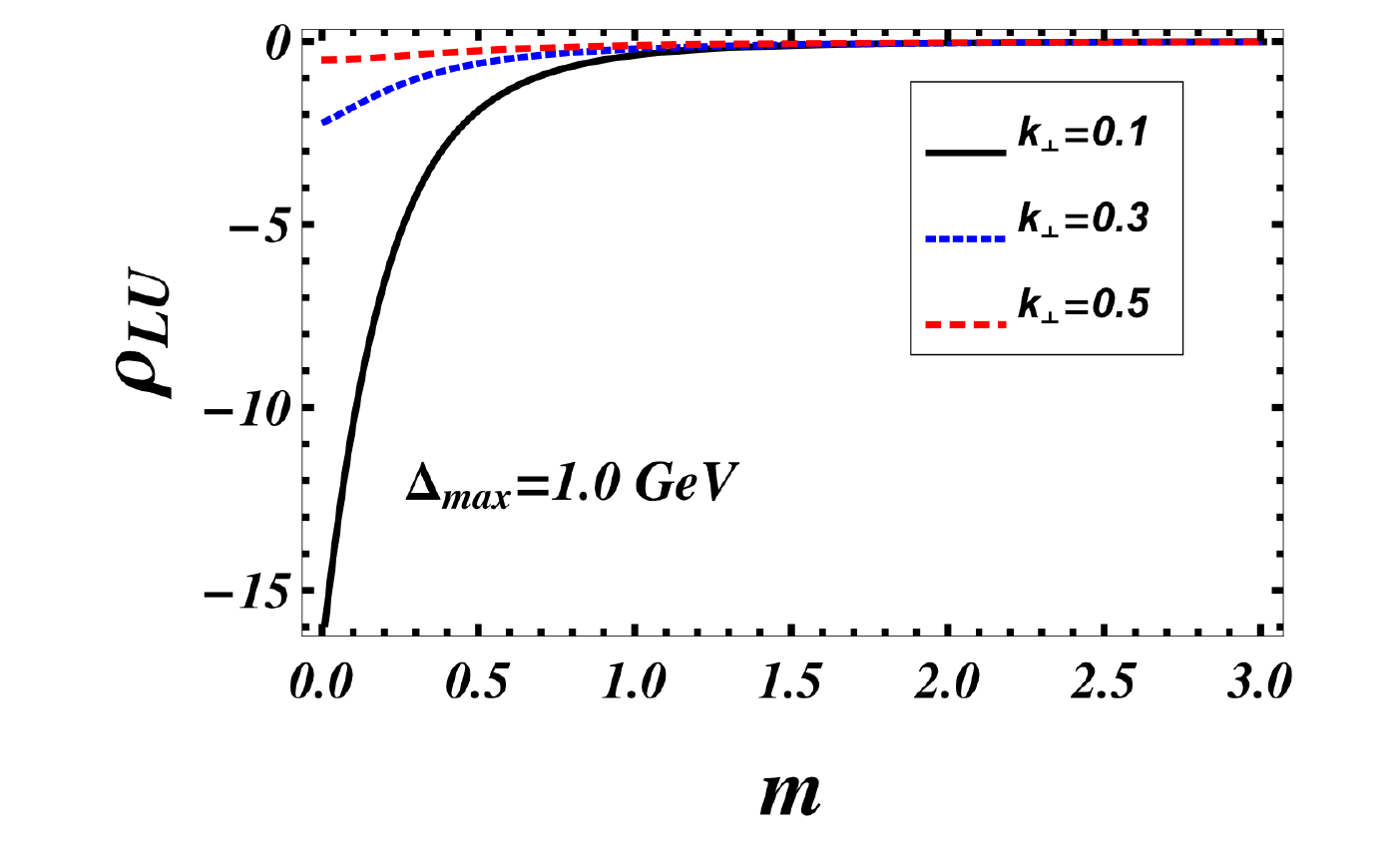}
\end{minipage}
\begin{minipage}[c]{1\textwidth}
\tiny{(e)}\includegraphics[width=7.8cm,height=5cm,clip]{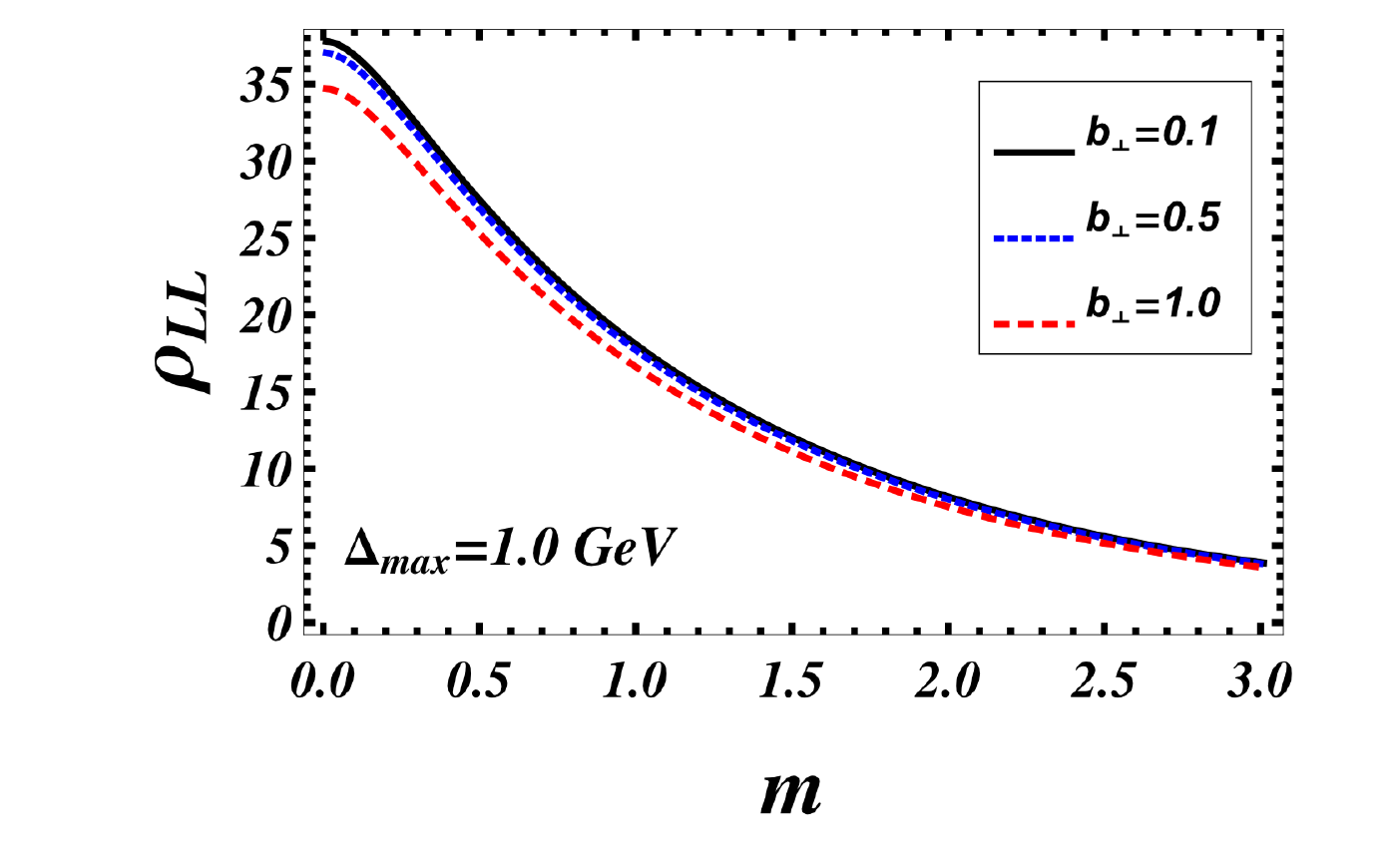}
\hspace{0.1cm}
\tiny{(f)}\includegraphics[width=7.8cm,height=5cm,clip]{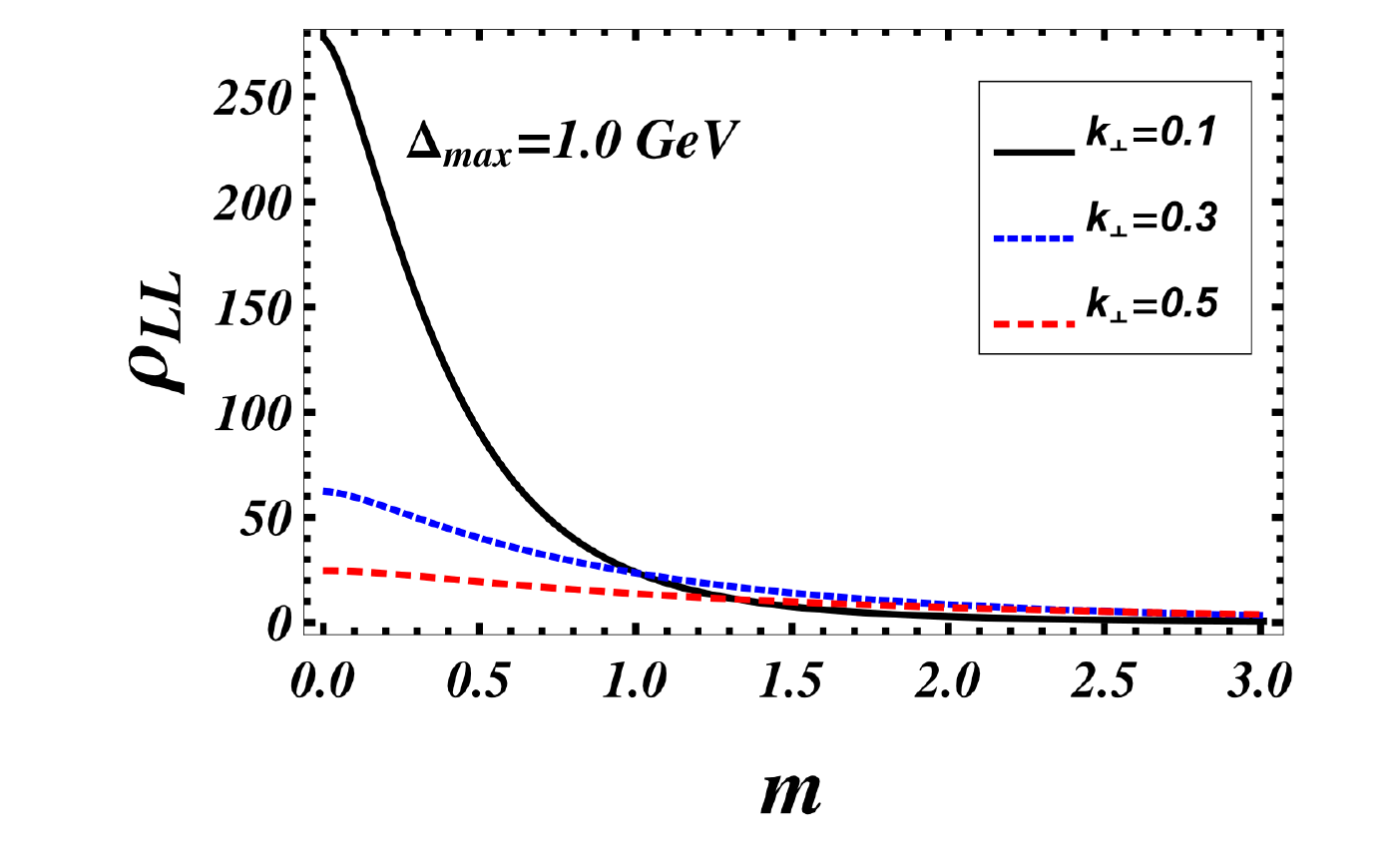}
\end{minipage}
\caption{\label{fig1}(Color online)
Plots of the Wigner distributions vs m (mass in $GeV$) for fixed values of $b_\perp$ and $k_\perp$ at
$\Delta_{max} = 1.0$ GeV. All the plots on the left (a,c,e) are for three fixed values of $b_\perp$ (0.1,0.5,1.0) in $GeV^{-1}$ where
 $k_\perp=0.4$ GeV. Plots on the right (b,d,f) are for three fixed values of $k_\perp$ (0.1,0.3,0.5)  in $GeV$ where
 and $b_\perp=0.4$ $GeV^{-1}$ .  For all plots we took $\vec{k_\perp} = k \hat{j}$ and $\vec{b_\perp} = b
\hat{j}$.
}
\end{figure}

\begin{figure}[!htp]
\begin{minipage}[c]{1\textwidth}
\tiny{(a)}\includegraphics[width=7.8cm,height=6cm,clip]{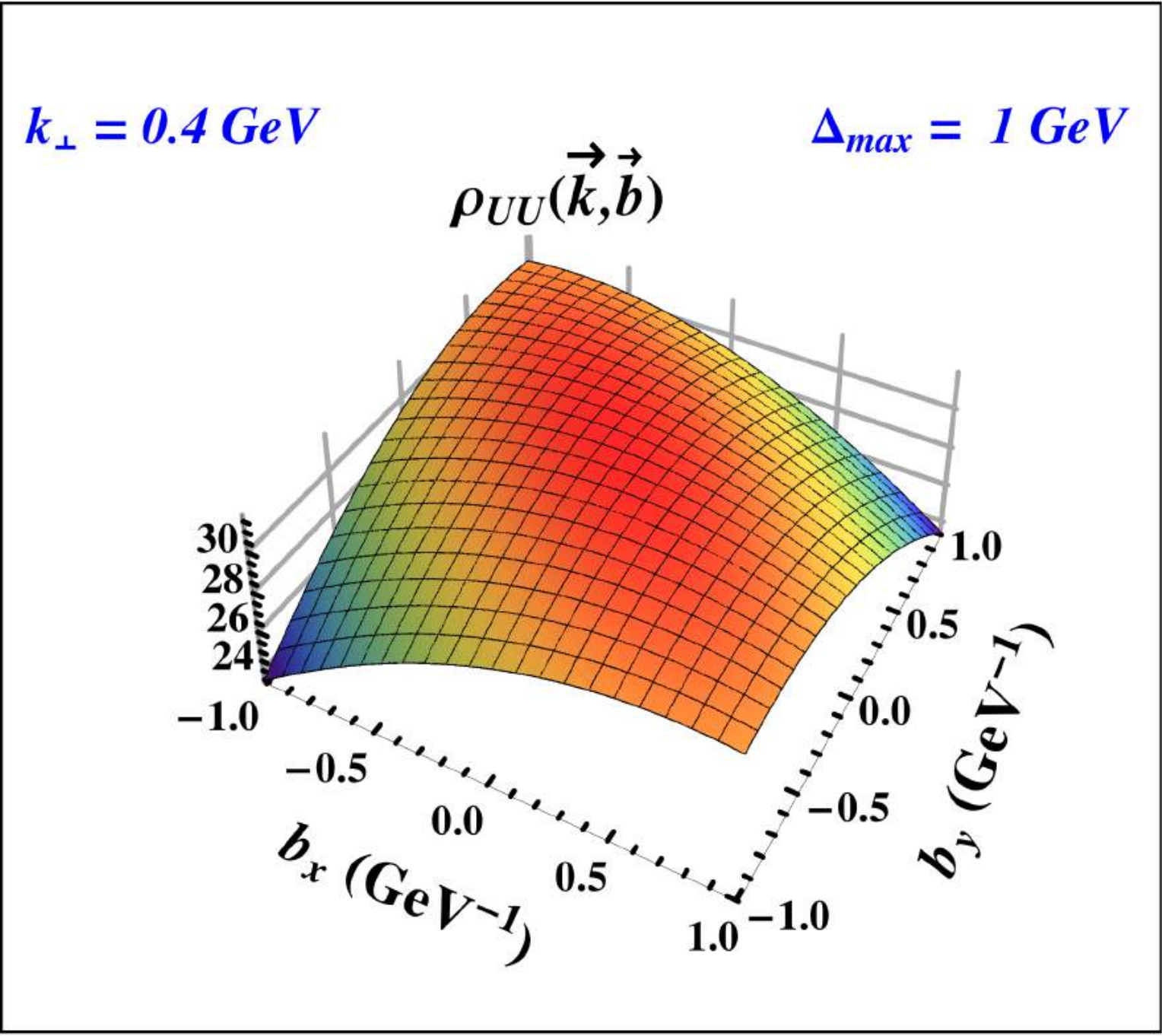}
\hspace{0.1cm}
\tiny{(b)}\includegraphics[width=7.8cm,height=6cm,clip]{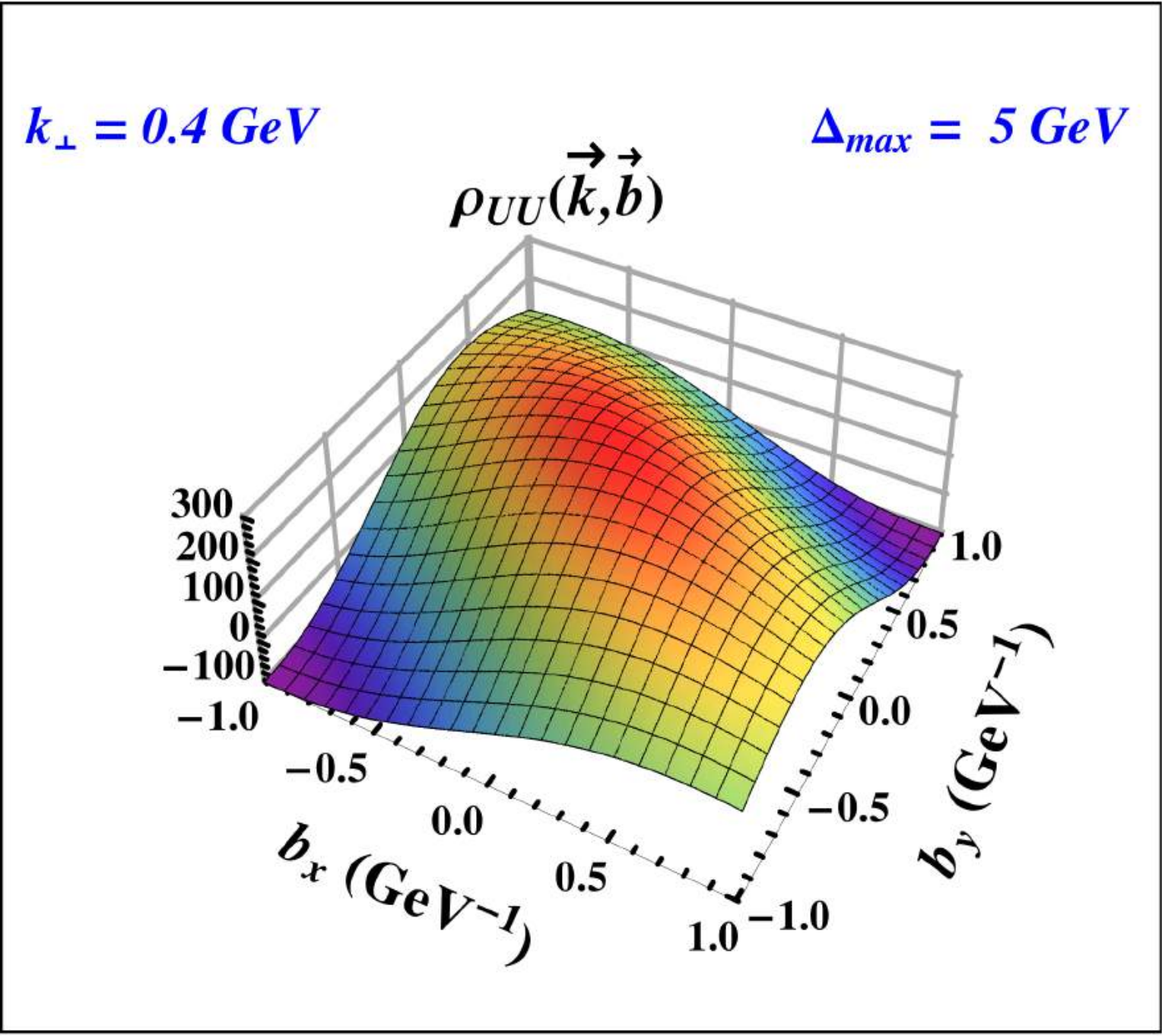}
\end{minipage}
\begin{minipage}[c]{1\textwidth}
\tiny{(c)}\includegraphics[width=7.8cm,height=6cm,clip]{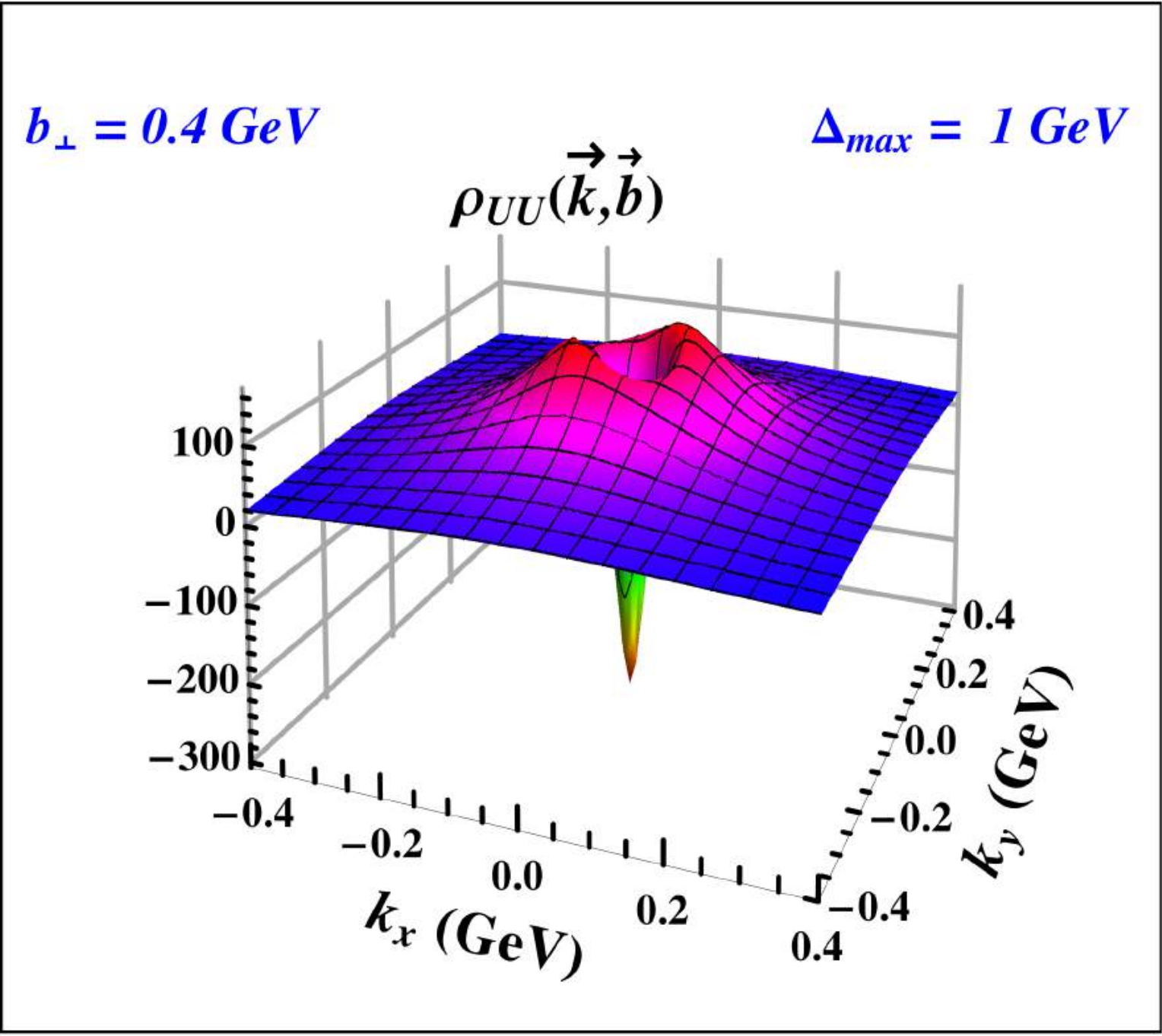}
\hspace{0.1cm}
\tiny{(d)}\includegraphics[width=7.8cm,height=6cm,clip]{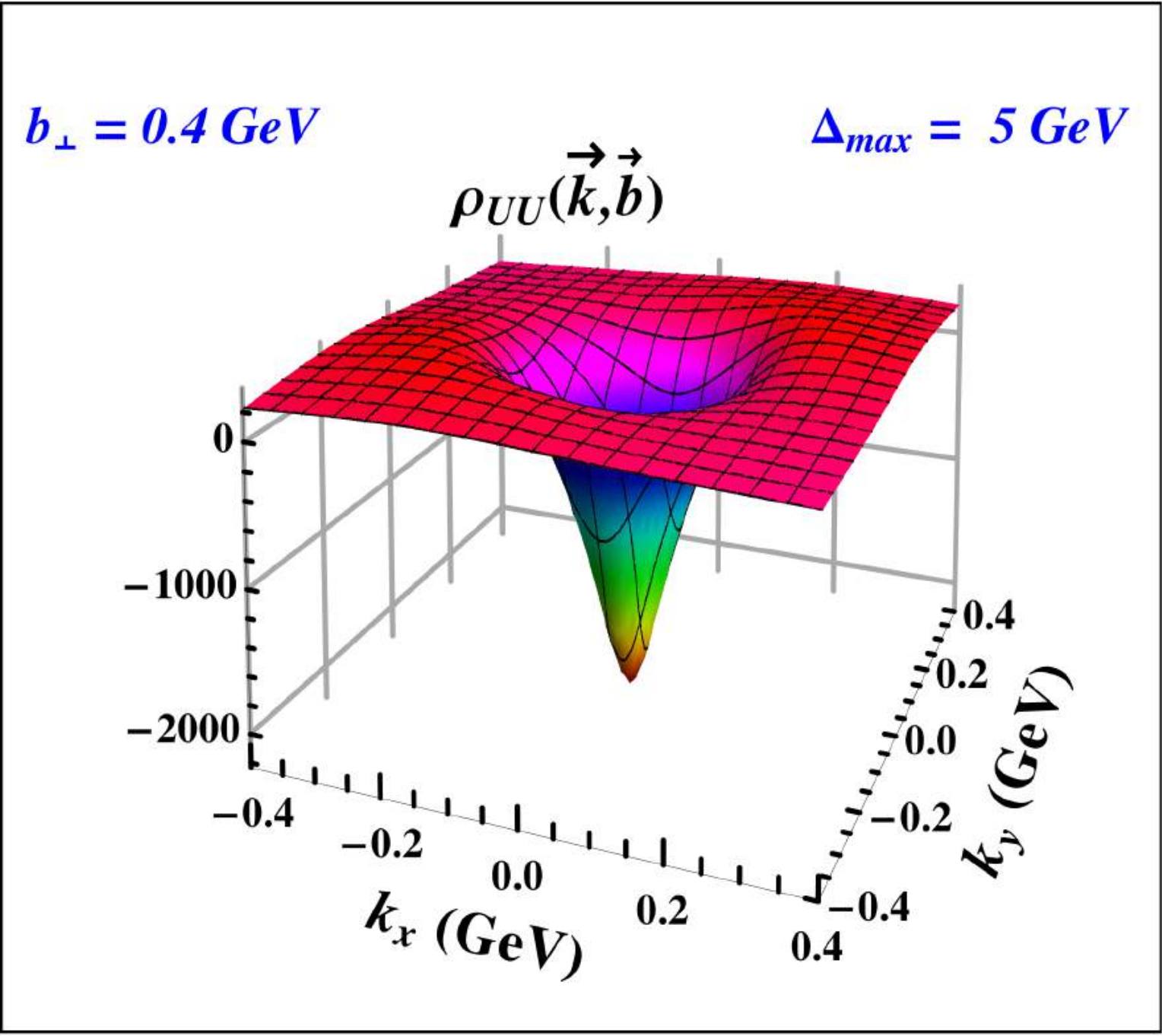}
\end{minipage}
\begin{minipage}[c]{1\textwidth}
\tiny{(e)}\includegraphics[width=7.8cm,height=6cm,clip]{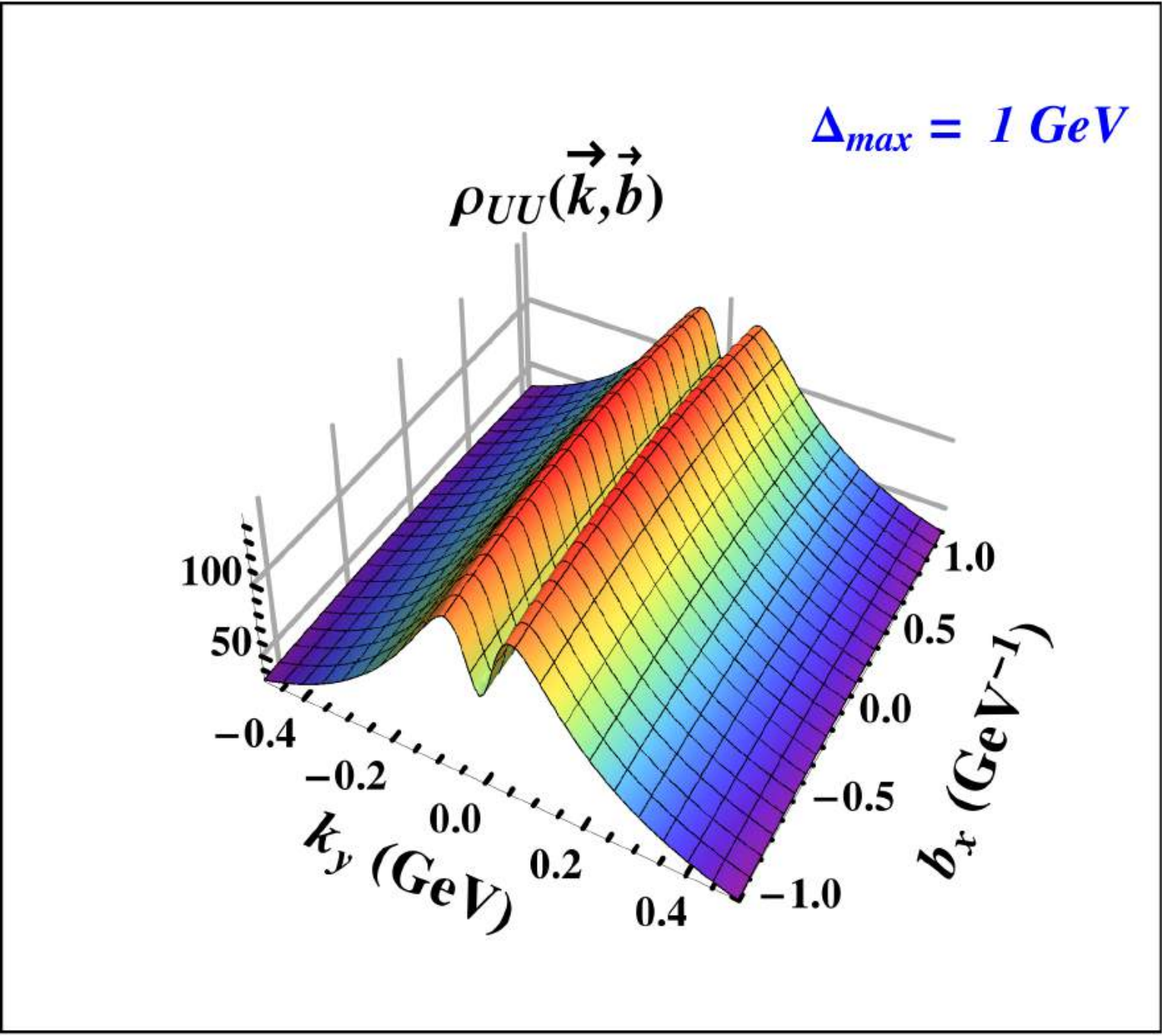}
\hspace{0.1cm}
\tiny{(f)}\includegraphics[width=7.8cm,height=6cm,clip]{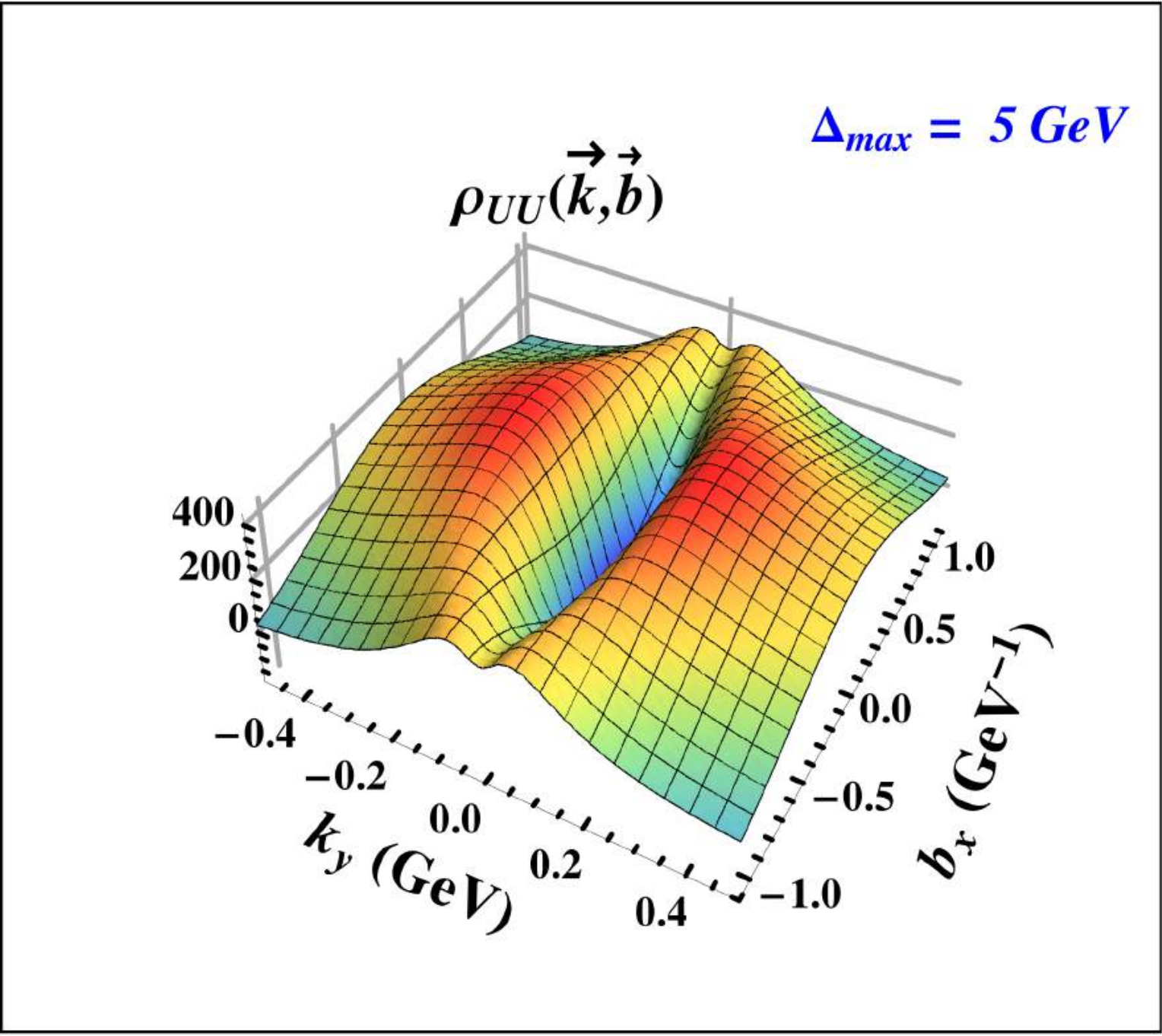}
\end{minipage}
\caption{\label{fig2}(Color online)
3D plots of the Wigner distributions $\rho_{UU}$. Plots (a) and (b) are in $b$ space with $k_\perp = 0.4$ GeV.
Plots (c) and (d) are in $k$ space with $b_\perp = 0.4$ $GeV^{-1}$.
Plots (e) and (f) are in mixed space where $k_x$ and $b_y$ are integrated.
All the plots on the left panel (a,c,e) are for $\Delta_{max} = 1.0$ GeV. Plots on the right panel (b,d,f) are for $\Delta_{max} = 5.0$ GeV.
For all the plots we kept $m = 0.33$ GeV, integrated out the $x$ variable and we took $\vec{k_\perp} = k \hat{j}$ and 
$\vec{b_\perp} = b \hat{j}$.  }
\end{figure}

\begin{figure}[!htp]
\begin{minipage}[c]{1\textwidth}
\tiny{(a)}\includegraphics[width=7.8cm,height=6cm,clip]{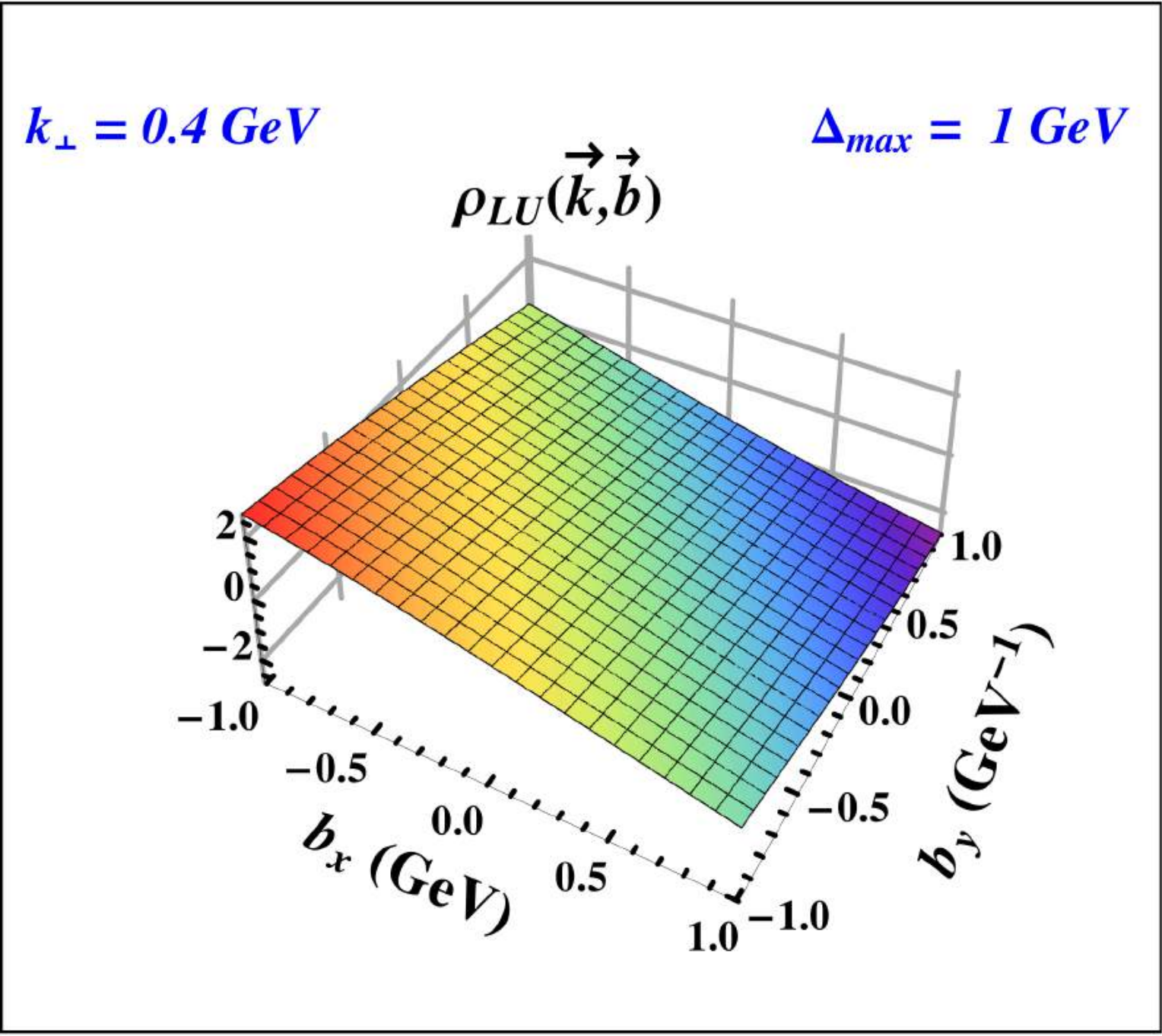}
\hspace{0.1cm}
\tiny{(b)}\includegraphics[width=7.8cm,height=6cm,clip]{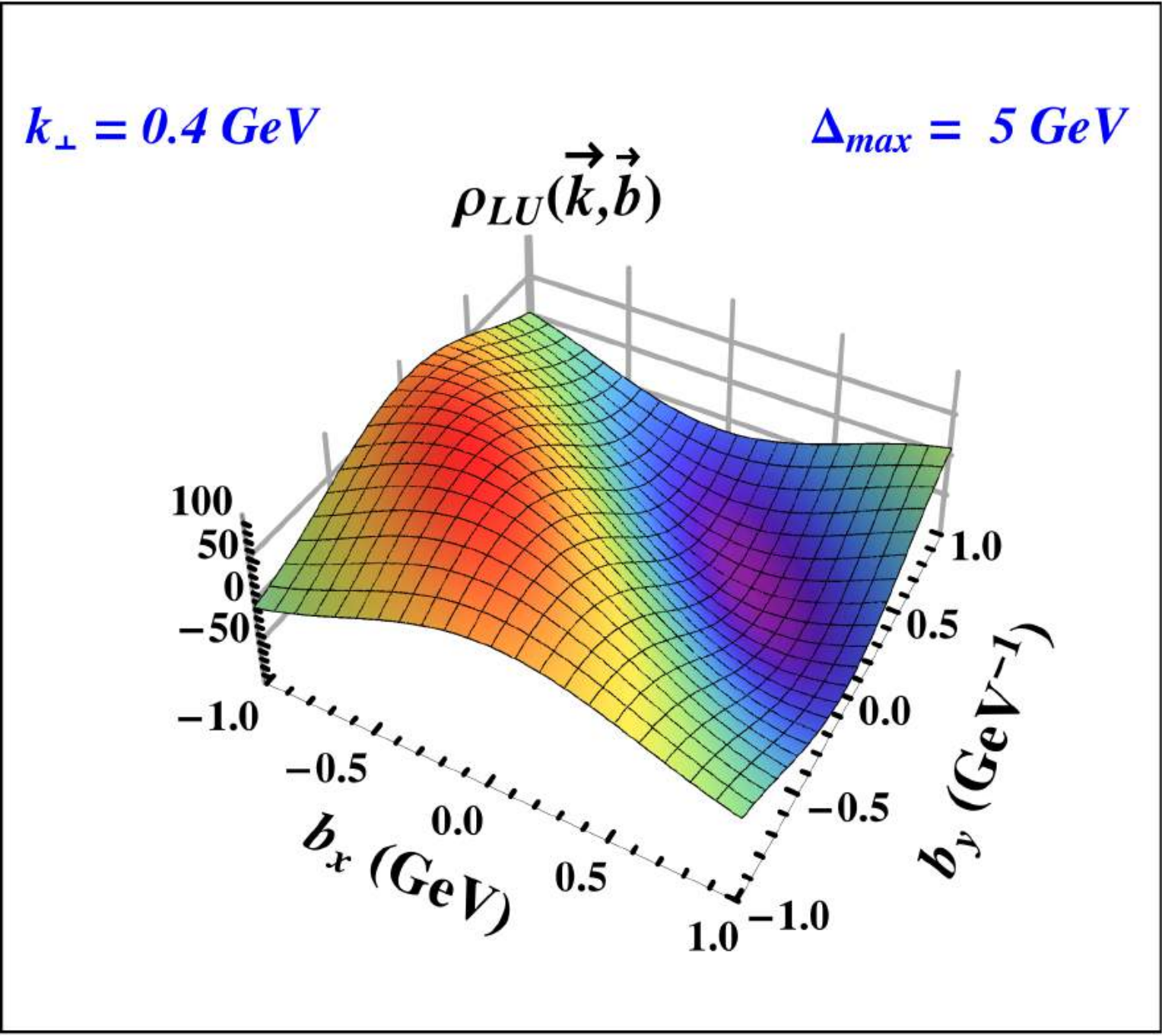}
\end{minipage}
\begin{minipage}[c]{1\textwidth}
\tiny{(c)}\includegraphics[width=7.8cm,height=6cm,clip]{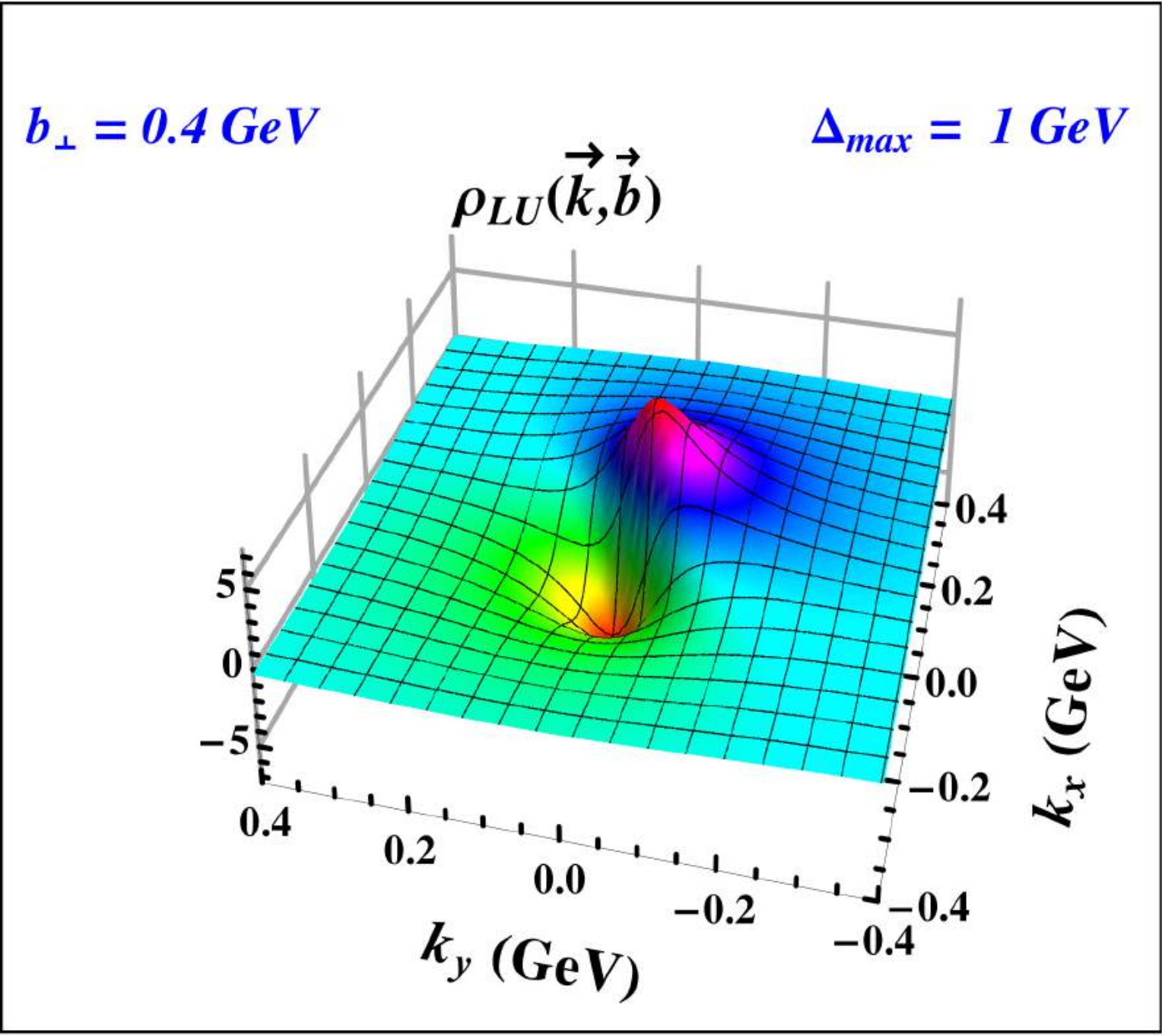}
\hspace{0.1cm}
\tiny{(d)}\includegraphics[width=7.8cm,height=6cm,clip]{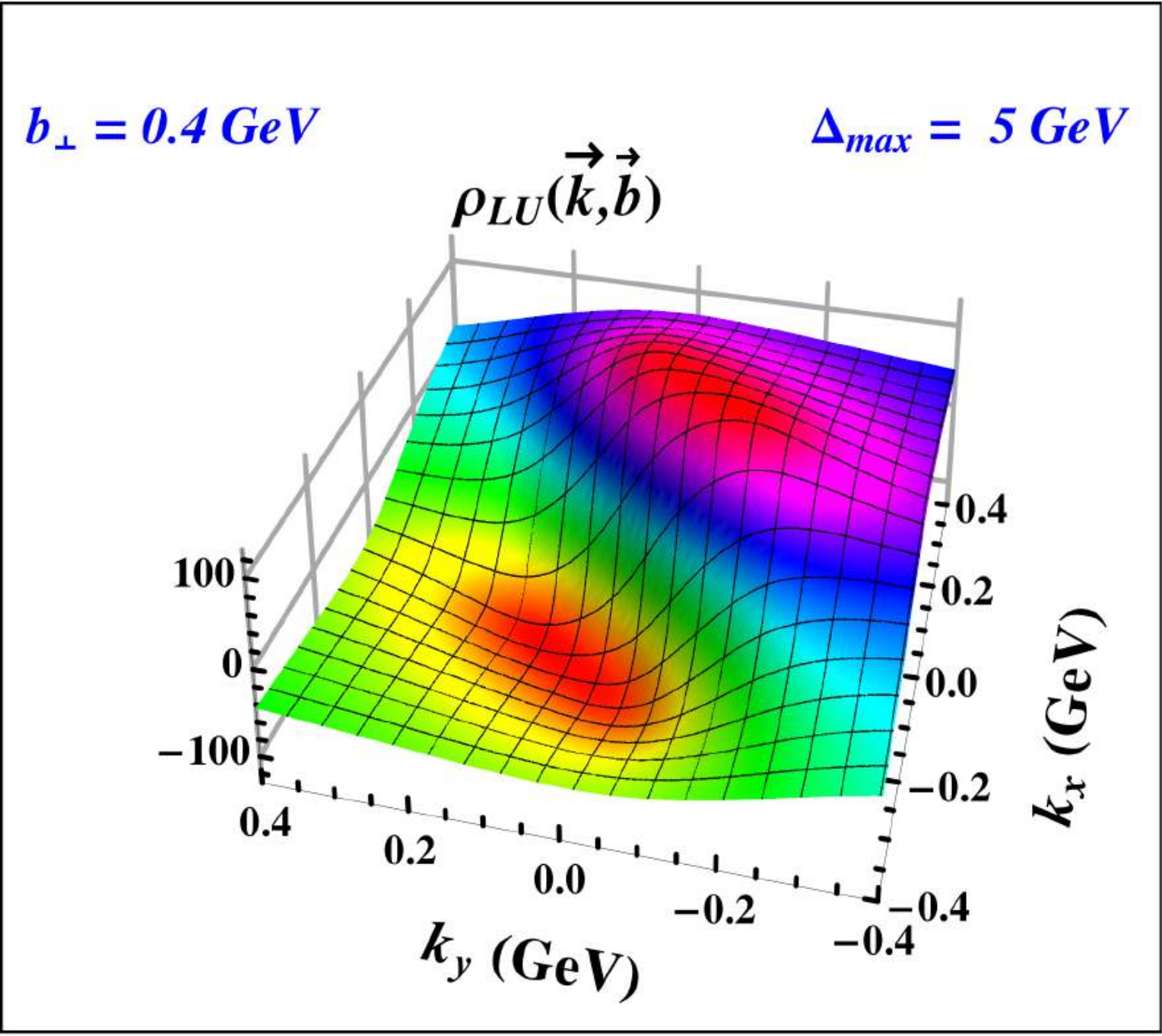}
\end{minipage}
\begin{minipage}[c]{1\textwidth}
\tiny{(e)}\includegraphics[width=7.8cm,height=6cm,clip]{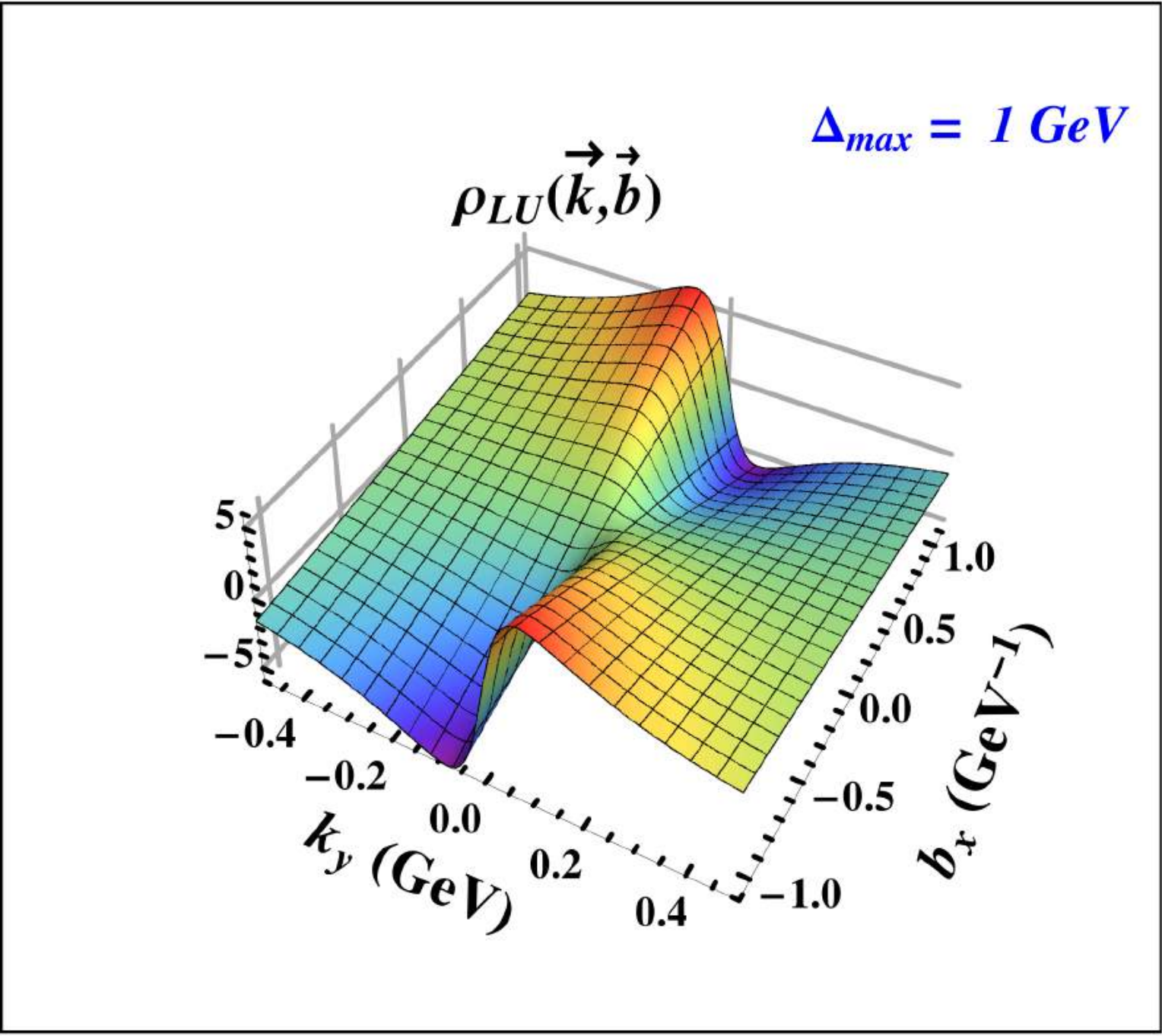}
\hspace{0.1cm}
\tiny{(f)}\includegraphics[width=7.8cm,height=6cm,clip]{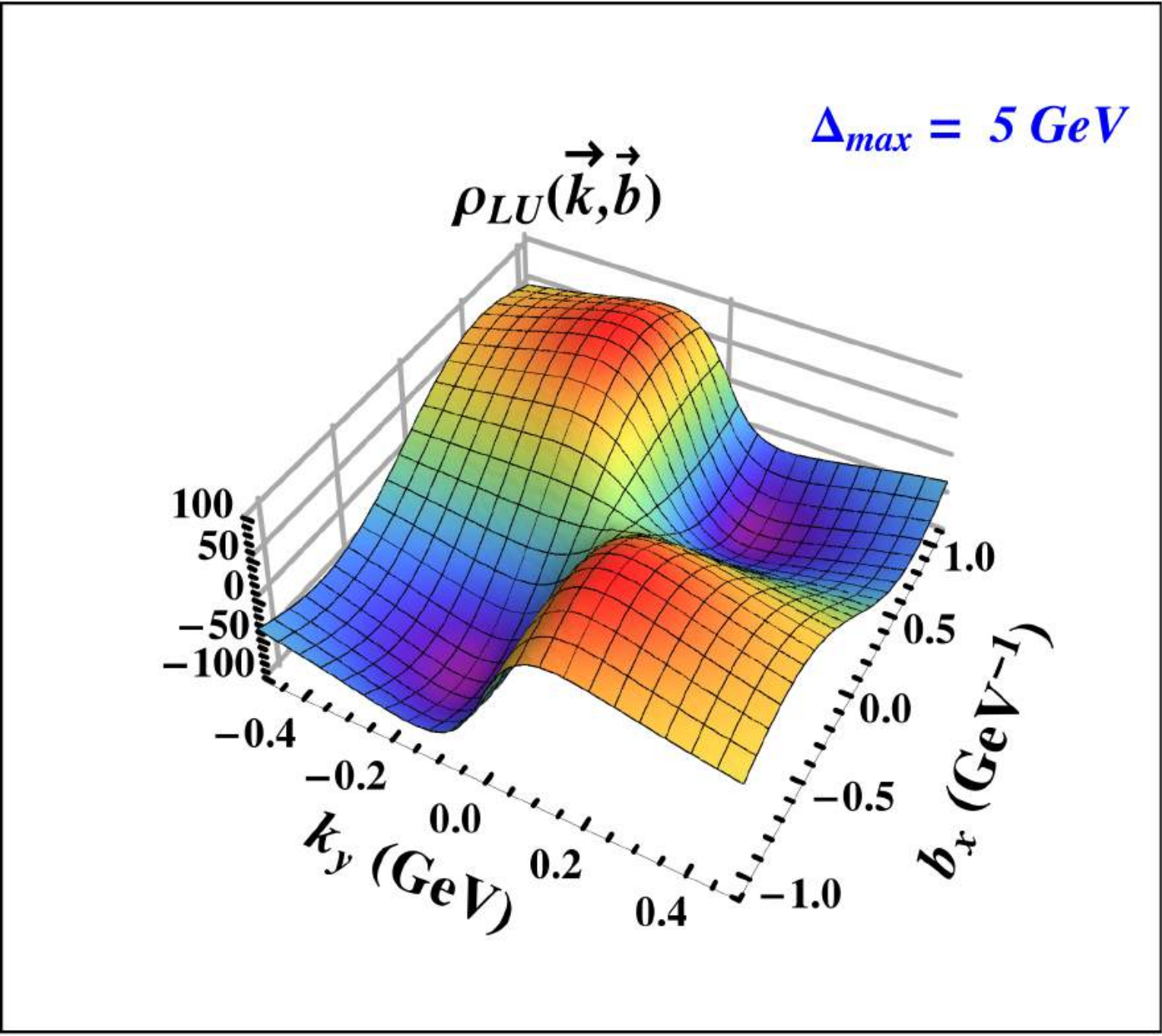}
\end{minipage}
\caption{\label{fig3}(Color online)
3D plots of the Wigner distributions $\rho_{LU}$. Plots (a) and (b) are in $b$ space with $k_\perp = 0.4$ GeV.
Plots (c) and (d) are in $k$ space with $b_\perp = 0.4$ $GeV^{-1}$.
Plots (e) and (f) are in mixed space where $k_x$ and $b_y$ are integrated.
All the plots on the left panel (a,c,e) are for $\Delta_{max} = 1.0$ GeV. Plots on the right panel (b,d,f) are for $\Delta_{max} = 5.0$ GeV.
For all the plots we kept $m = 0.33$ GeV, integrated out the $x$ variable  and we took $\vec{k_\perp} = k \hat{j}$ 
and $\vec{b_\perp} = b \hat{j}$.  }
\end{figure}

\begin{figure}[!htp]
\begin{minipage}[c]{1\textwidth}
\tiny{(a)}\includegraphics[width=7.8cm,height=6cm,clip]{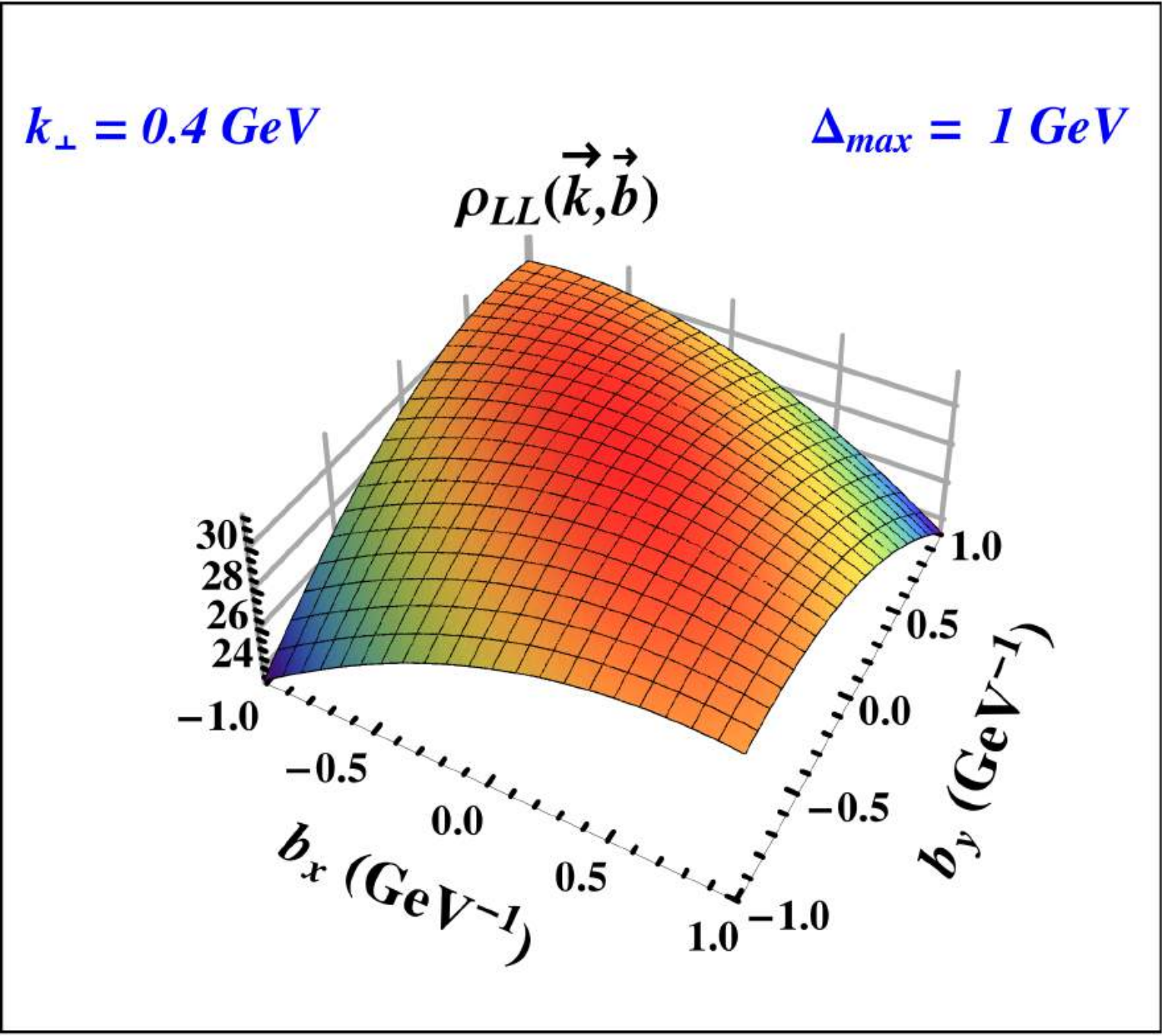}
\hspace{0.1cm}
\tiny{(b)}\includegraphics[width=7.8cm,height=6cm,clip]{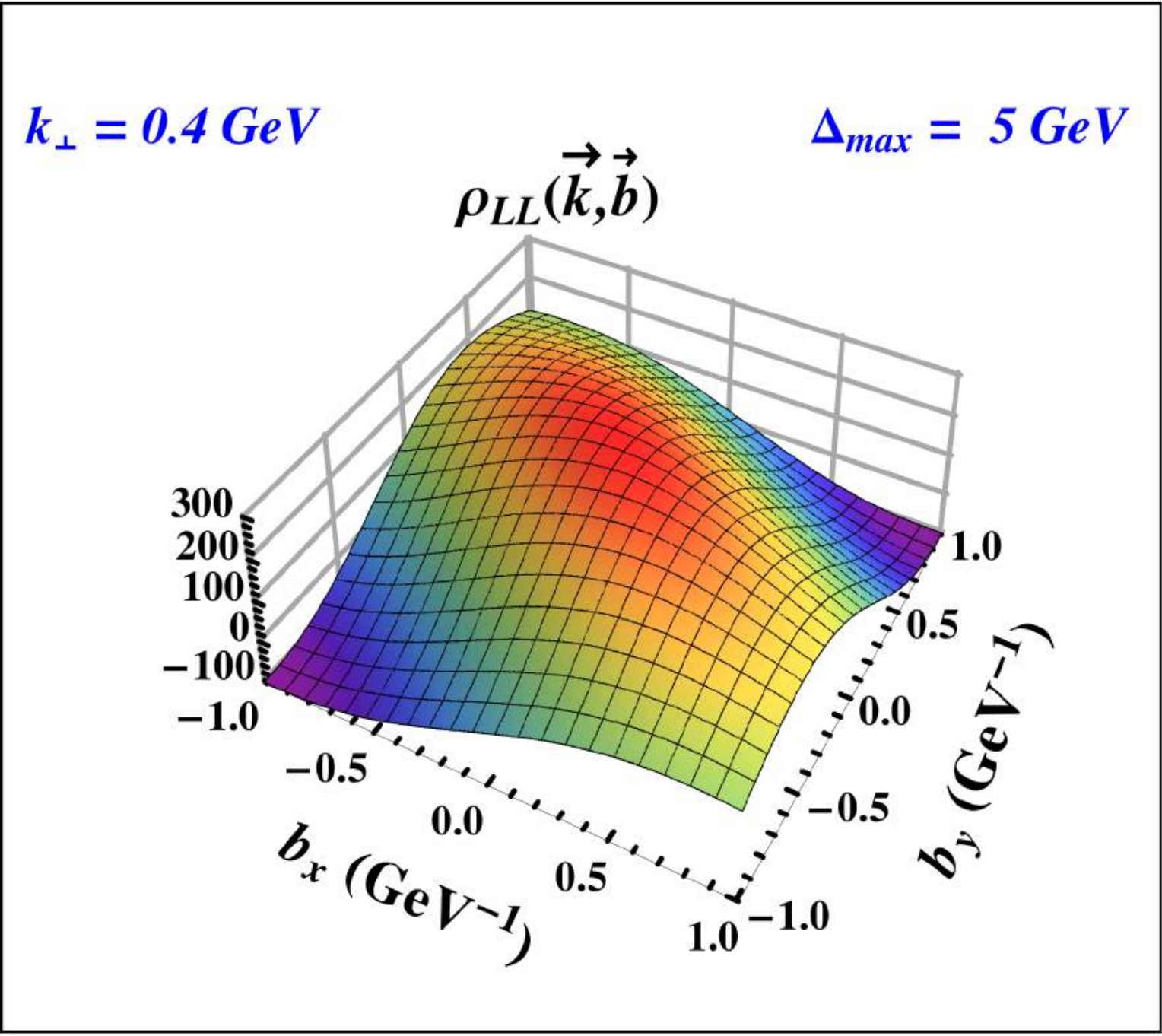}
\end{minipage}
\begin{minipage}[c]{1\textwidth}
\tiny{(c)}\includegraphics[width=7.8cm,height=6cm,clip]{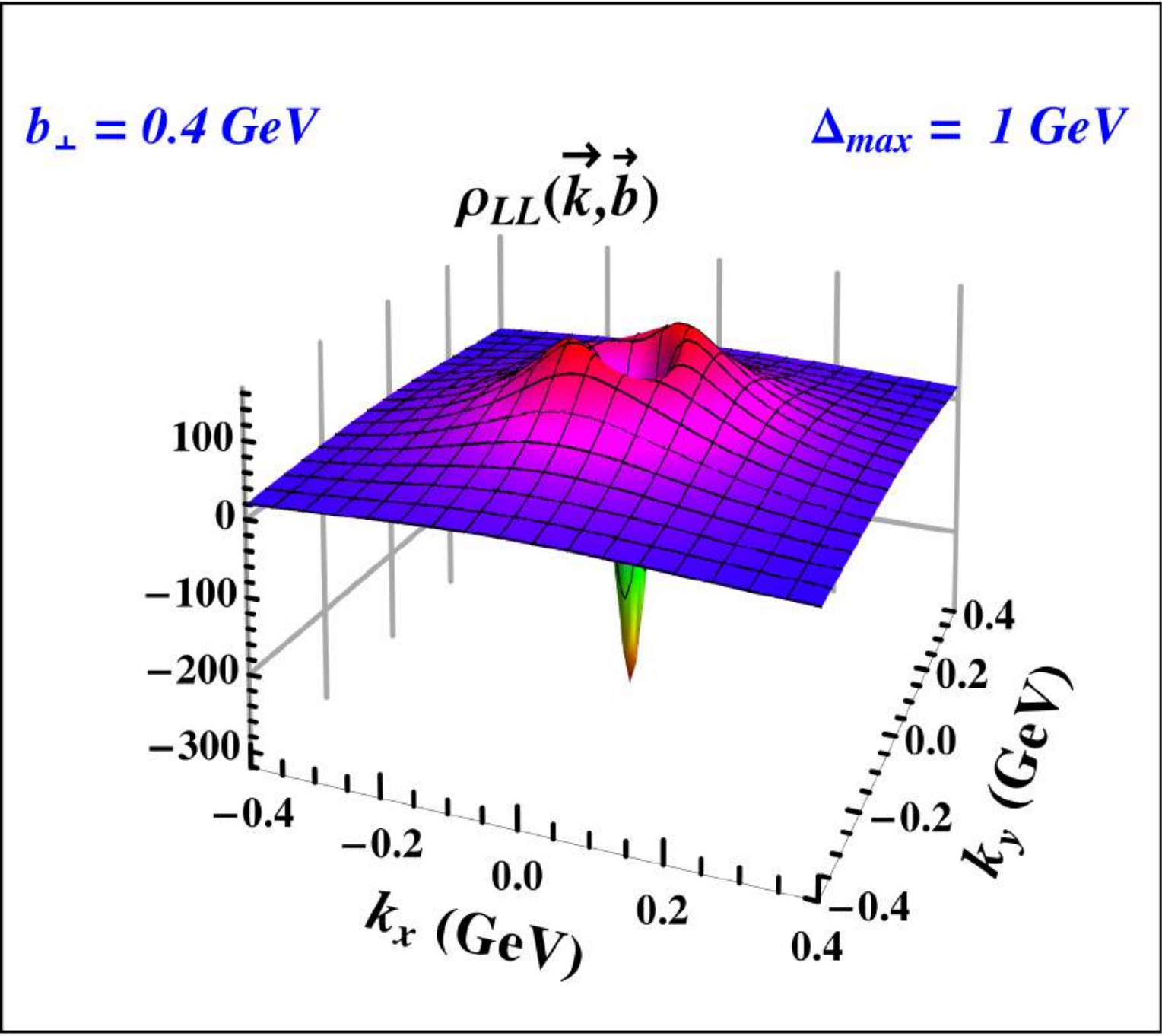}
\hspace{0.1cm}
\tiny{(d)}\includegraphics[width=7.8cm,height=6cm,clip]{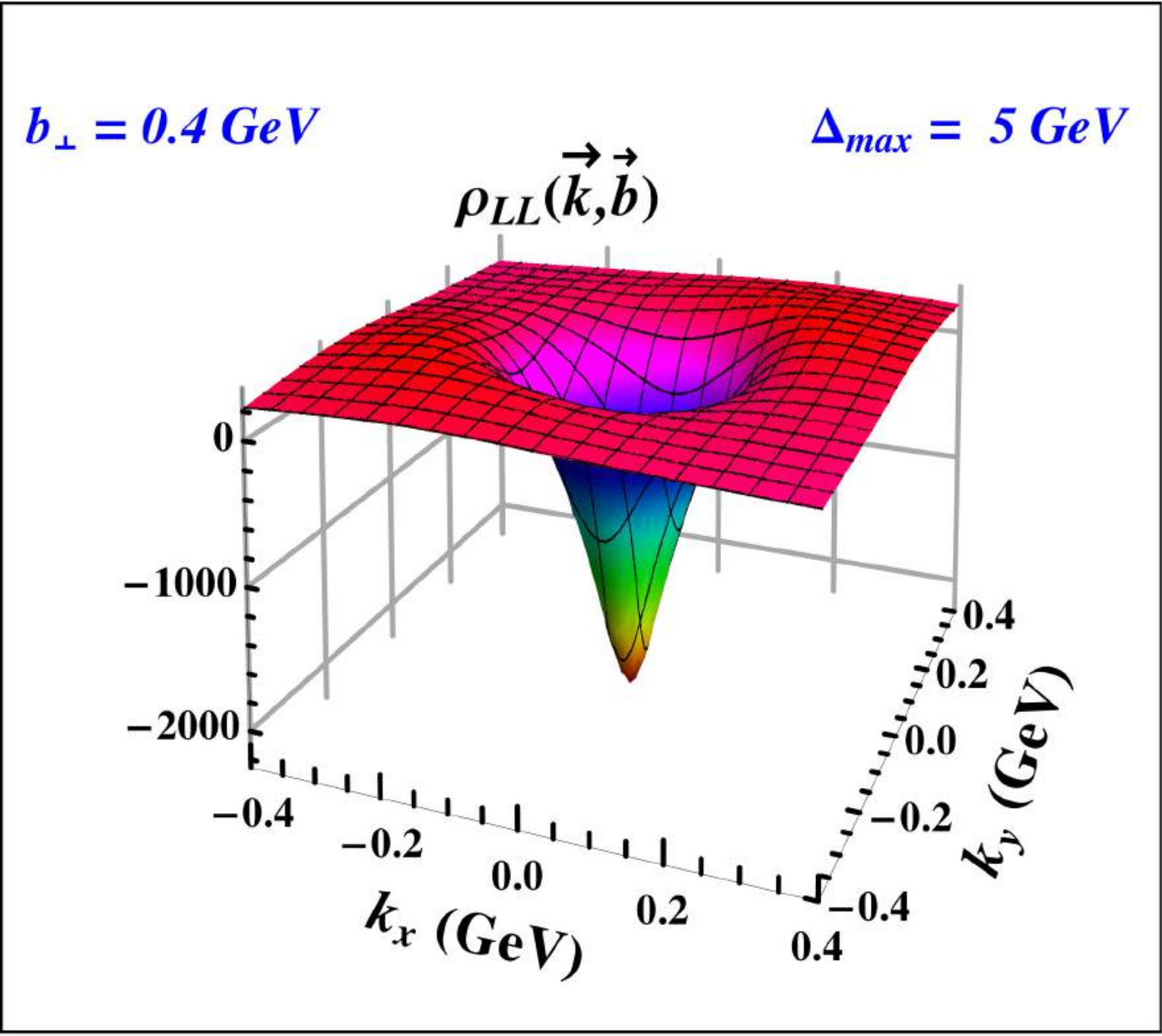}
\end{minipage}
\begin{minipage}[c]{1\textwidth}
\tiny{(e)}\includegraphics[width=7.8cm,height=6cm,clip]{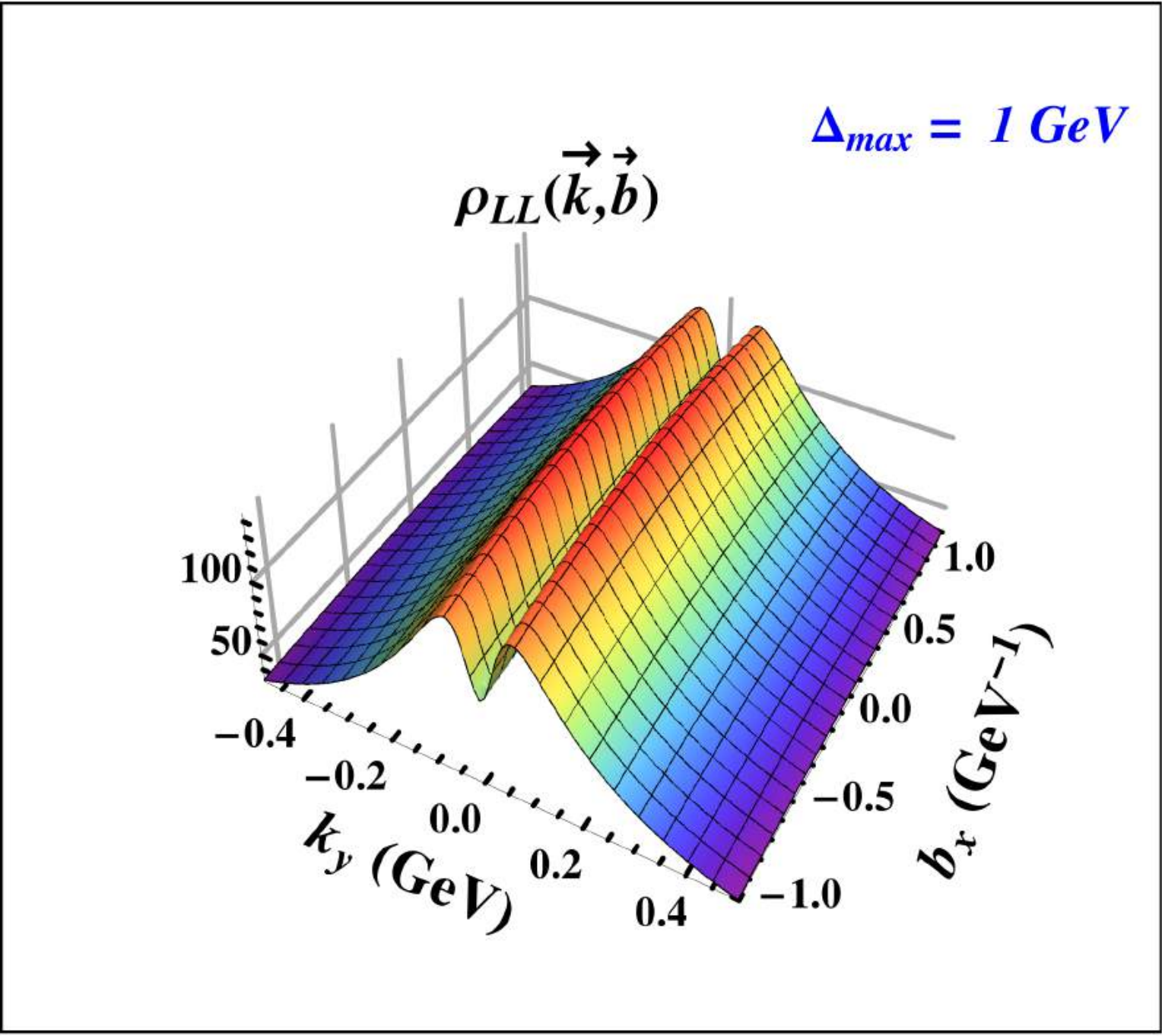}
\hspace{0.1cm}
\tiny{(f)}\includegraphics[width=7.8cm,height=6cm,clip]{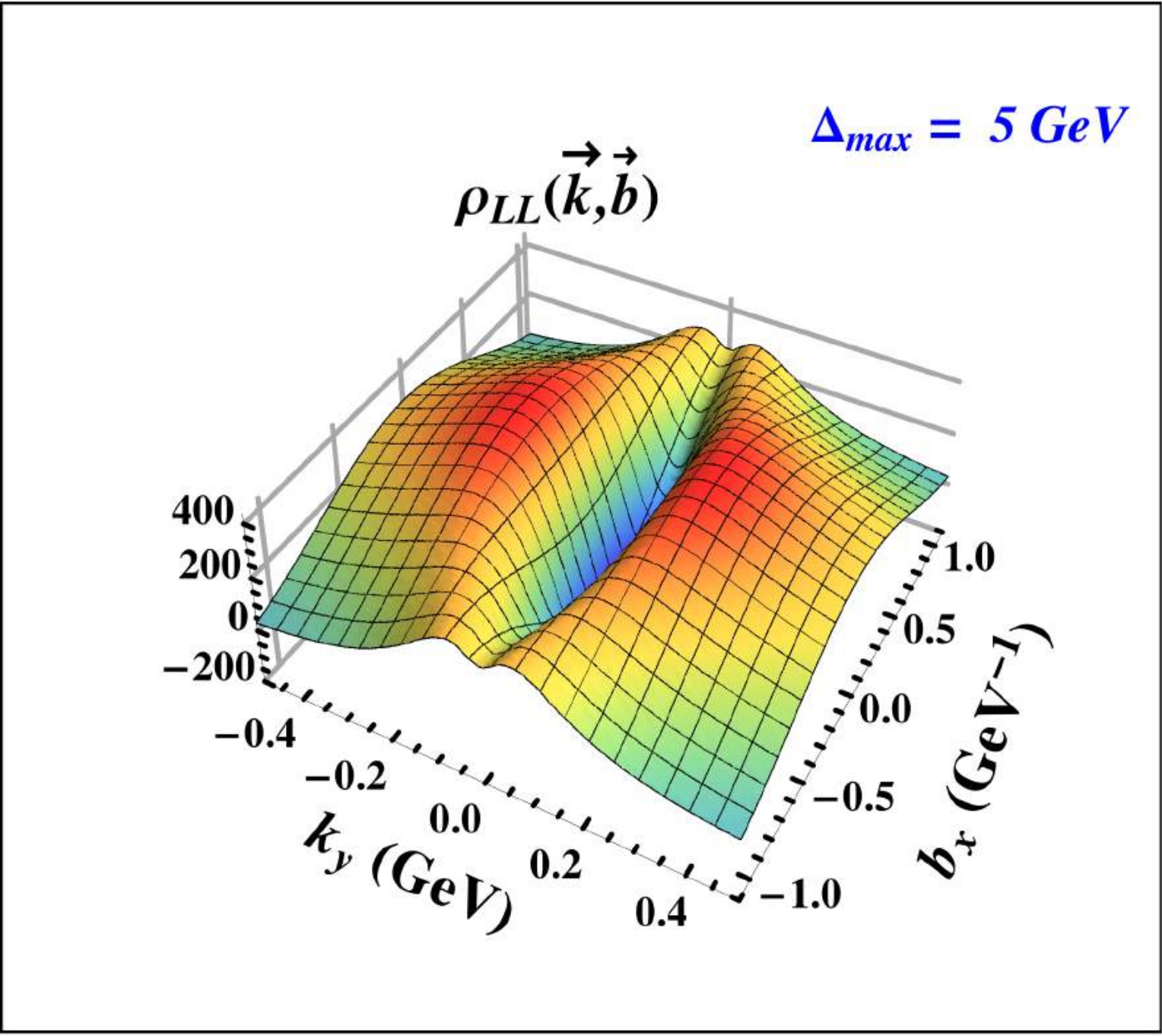}
\end{minipage}
\caption{\label{fig4}(Color online)
3D plots of the Wigner distributions $\rho_{LL}$. Plots (a) and (b) are in $b$ space with $k_\perp = 0.4$ GeV.
Plots (c) and (d) are in $k$ space with $b_\perp = 0.4$ $ GeV^{-1}$.
Plots (e) and (f) are in mixed space where $k_x$ and $b_y$ are integrated.
All the plots on the left panel (a,c,e) are for $\Delta_{max} = 1.0 GeV$. Plots on the right panel (b,d,f) are for $\Delta_{max} = 5.0$ GeV.
For all the plots we kept $m = 0.33$ GeV, integrated out the $x$ variable and we took $\vec{k_\perp} = k \hat{j}$ 
and $\vec{b_\perp} = b \hat{j}$.  }
\end{figure}

\section{Orbital Angular Momentum of quarks}
\noindent
In \cite{metz} it has been shown that the quark-quark correlator in Eq.(\ref{qqc}) defining the Wigner distributions  
can be parameterized in terms of generalized transverse momentum dependent parton distributions (GTMDs).
For the twist two case we have four GTMDs $(F_{1,i})$  corresponding to $\gamma^+$  
and four more for $\gamma^+\gamma_5$ $ (G_{1,i})$

\be
W_{\lambda,\lambda'}^{[\gamma^+]} =
\frac{1}{2M} \bar{u}(p',\lambda') \Big[
F_{1,1}-
\frac{i \sigma^{i+}k_{i\perp}}{P^+} F_{1,2} -
 \frac{i\sigma^{i+}\Delta_{i \perp}}{P^+} F_{1,3} +
\frac{i\sigma^{ij}k_{i\perp}\Delta_{j\perp}}{M^2} F_{1,4}
\Big]
u(p,\lambda);
\label{f11}\ee \\

\be
W_{\lambda,\lambda'}^{[\gamma^+\gamma_5]} =
\frac{\bar{u}(p',\lambda') }{2M} \Big[
\frac{-i \epsilon^{ij}_{\perp} k_{i\perp}\Delta_{j\perp}}{M^2}G_{1,1}-
\frac{i \sigma^{i+}\gamma_{5} k_{i\perp}}{P^+}G_{1,2} -
 \frac{i\sigma^{i+}\gamma_{5} \Delta_{i\perp}}{P^+} G_{1,3} +
i\sigma^{+-} \gamma_5 G_{1,4}
\Big]
u(p,\lambda).
\label{g11}\nn \ee \\%
\noindent
Using the above two equations and Eq.(\ref{main}) we calculate the GTMDs for the dressed quark model at 
twist two.
We have used the Bjorken and Drell convention for gamma matrices. Using the
two-particle LFWFs we obtain the final expression for the GTMDS as follows :

\begin{align} 
F_{11} &= - \frac{N \Big[4k_{\perp}^{2}(1+x^2) +(x-1)^2 (4m^2(x-1)^2 -(1+x^2) \Delta_{\perp}^{2}) \Big]}{D(q_{\perp})
D(q'_{\perp})2x^2(x-1)^3};  \\
\nn \\
 F_{12} &= \frac{2N m^2  \Delta_{\perp}^{2}}{D(q_{\perp})D(q'_{\perp})x(k_y \Delta_x - k_x
\Delta_y)}; \\ 
 \nn \\
 F_{13} & =  \frac{N}{D(q_{\perp})D(q'_{\perp})4x(k_y \Delta_x - k_x \Delta_y)}\nn \\ 
 \Big[ 
  & 8 m^2(k_{\perp} \Delta_{\perp} ) - 
  \frac{(k_y \Delta_x - k_x \Delta_y)(4k_{\perp}^{2}(1+x^2) +(x-1)^2 (4 m^2(x-1)^2 -(1+x^2) \Delta_{\perp}^{2}) )}
  {x(x-1)^3}
  \Big];\nn  \\
   \\
 F_{14} &= \frac{2N m^2(1+x)}{D(q_{\perp})D(q'_{\perp})x^2(1-x)}.  
 \end{align}

\begin{align} 
 G_{11} &= -\frac{2N m^2(1+x)}{D(q_{\perp})D(q'_{\perp})x^2(x-1)};  \\
\nn \\
G_{12} &= \frac{-N }{D(q_{\perp})D(q'_{\perp})x(x-1)} \Big[  4m^2 \frac{k_\perp . \Delta_{\perp}}{(k_y \Delta_x - k_x \Delta_y)}
 -\frac{(1+x)\Delta_{\perp}^{2}}{x}
 \Big]; \\
 \nn \\
G_{13} &=  \frac{N \Big[ 
  (1+x)\Big( \Delta_y^{2} -   \Delta_x^{2}  + \Delta_x \Delta_y (k_y^{2} -
k_x^{2})  \Big) + 4x
m^2 k_{\perp}^{2}
  \Big]  }{D(q_{\perp})D(q'_{\perp})x^2(x-1)(k_y \Delta_x - k_x \Delta_y)}; \\ 
  \nn \\
 G_{14} &= \frac{N \Big[ -4k_{\perp}^{2}(1+x^2) +(x-1)^2 \Big(4 m^2(x-1)^2 -(1+x^2) 
\Delta_{\perp}^{2}\Big) \Big]}{D(q_{\perp})D(q'_{\perp})2x^2(x-1)^3}; 
 \end{align} \\

 where $N = \frac{g^2 C_f }{2(2\pi)^3}$ is the normalization constant and
$C_f$  is the color factor.\\
 \\
\noindent
The kinetic quark orbital angular momentum (OAM) is given in terms of the
GPDs \cite{Ji} as:

\begin{figure}[!htp]
\begin{minipage}[c]{1\textwidth}
\tiny{(a)}\includegraphics[width=7.8cm,height=6cm,clip]{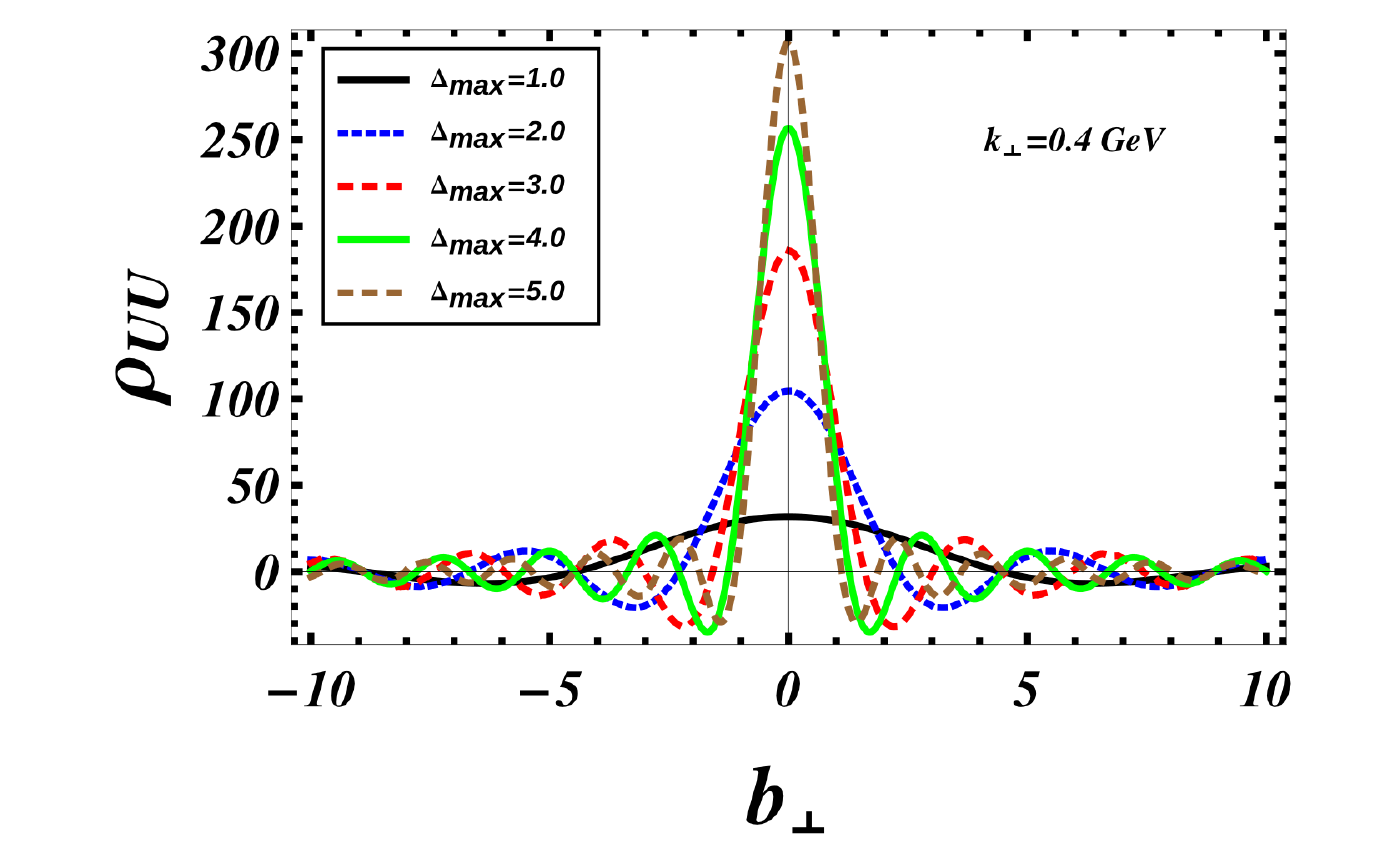}
\hspace{0.1cm}
\tiny{(b)}\includegraphics[width=7.8cm,height=6cm,clip]{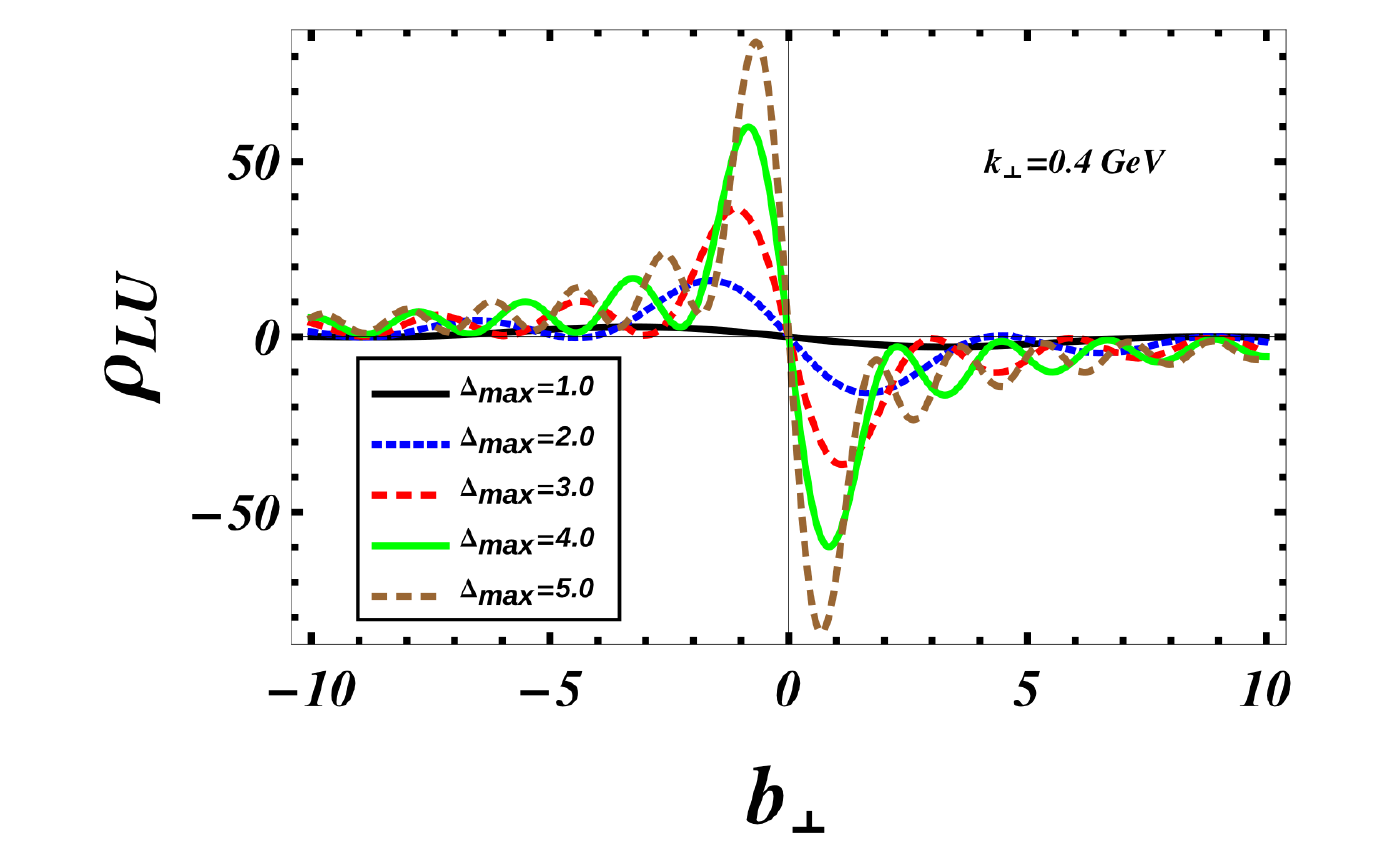}
\end{minipage}
\begin{minipage}[c]{1\textwidth}
\tiny{(c)}\includegraphics[width=7.8cm,height=6cm,clip]{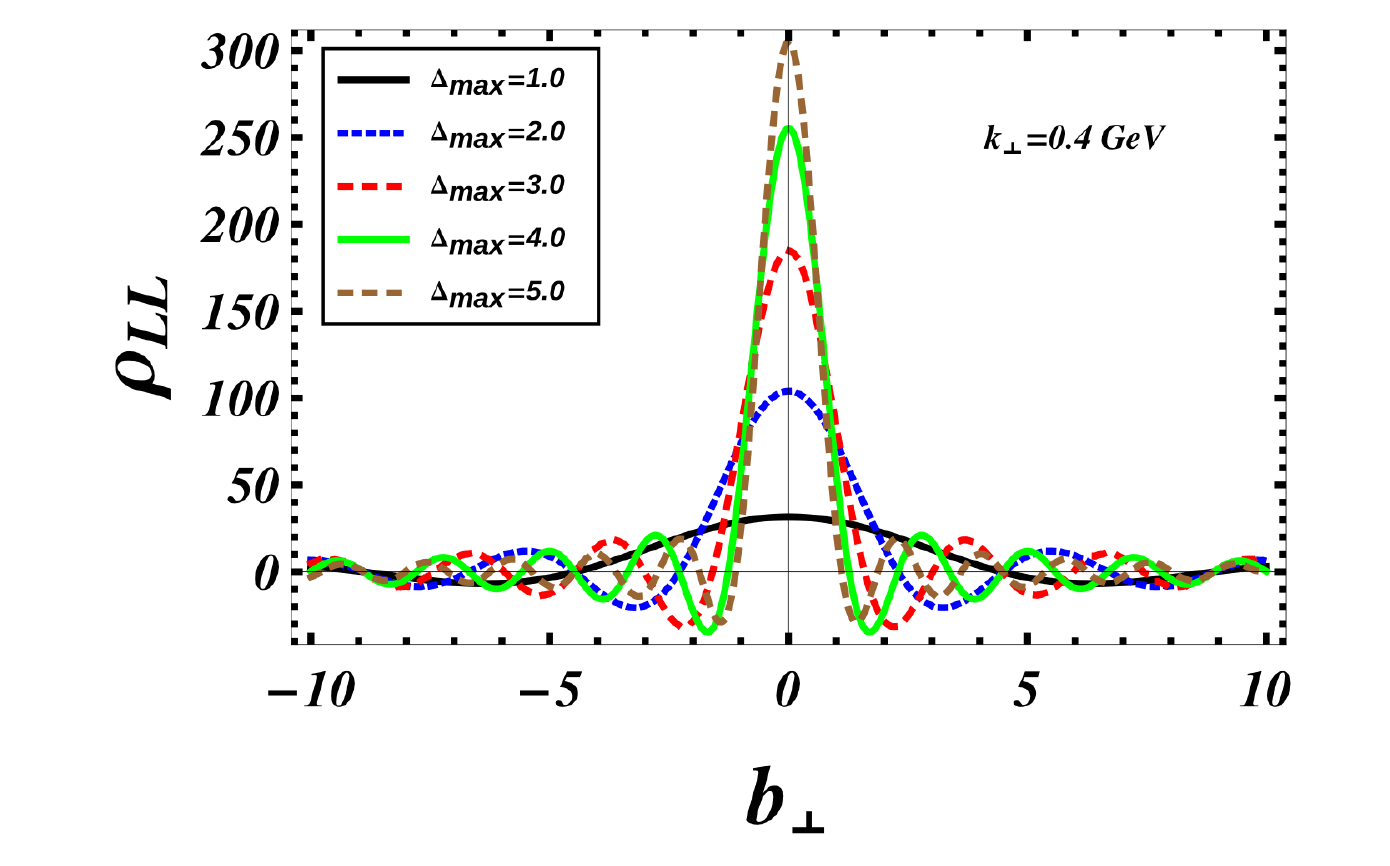}
\end{minipage}
\caption{\label{fig5}(Color online)
Plots of the Wigner distributions vs $ b_\perp $ for different 
 $\Delta_{max} (GeV)$ for a fixed value $k_\perp=0.4$ GeV and
m = $0.33$ GeV. $b_\perp$ is in $GeV^{-1}$. }
\end{figure}

\be 
L^{q}_{z} = \frac{1}{2} \int dx \{ 
x [ H^{q}(x,0,0) + E^{q}(x,0,0) ] - \tilde{H^q}(x,0,0)
\}. \nn
\ee
\\
\noindent
The GPDs in the above equation are defined at $\xi=0$ or when the momentum
transfer is purely in the transverse direction. GPDs in the model we
consider have been already calculated in
\cite{chak1},\cite{chak2},\cite{brod1},\cite{brod2},\cite{metz1}. 
The kinetic OAM is related to the GTMDs
\cite{metz} by the following relations:

\begin{align}
  H(x,0,t) &= \int d^{2} k_{\perp} F_{11};  \\
 E(x,0,t) &= \int d^{2} k_{\perp} \Big[ 
 -F_{11} +2\Big( 
 \frac{k_{\perp}.\Delta_{\perp}}{\Delta_{\perp}^{2}} F_{12} + F_{13}
 \Big)
 \Big];  \\
 \tilde{H}(x,0,t) &= \int d^{2} k_{\perp} G_{14}.
  \end{align} \\

\noindent
Using the GTMDs calculated we have the following final expression for the kinetic orbital angular 
momentum of quarks in the dressed quark model:\\
 
\be \label{cl}
 L^{q}_{z} = \frac{N}{2} \int dx \Big \{ 
 - f(x) I_{1}  +
4m^2(1-x)^2  I_{2}
\Big
\};
\ee 
 where,
 
 \begin{align}
  I_{1} &=   \int 
 \frac{d^{2} k_{\perp}}{m^2(1-x)^2 +(k_{\perp})^2}  = \pi log\Bigg[\frac{Q^2+m^2(1-x)^2}{\mu^2+m^2(1
-x)^2}\Bigg];\nn \\
  I_{2} &=  \int  \frac{d^{2} k_{\perp}}{\Big(m^2(1-x)^2 +(k_{\perp})^2\Big)^2} =
  \frac{\pi}{(m^2(1-x)^2)};\nn\\
  f(x) &=2(1+x^2). 
 \nn 
 \end{align}
Here $Q$ and $\mu$ are the upper and lower limits of the $k_\perp$ integration
respectively. $Q$ is the large scale involved in the process, which comes
from the large momentum cutoff in this approach \cite{hari}. Alternatively one
can choose an invariant mass cutoff \cite{brodsky_OAM}. 
$\mu$ can be safely taken to be zero provided the quark mass
is non-zero.  In fact, we have taken $\mu$ to be zero.\\%
\newline
The GTMDs $F_{14}$ and $G_{11}$ are not reducible to any GPDs or
transverse-momentum dependent parton distributions (TMDs) in any limit.
These  appear purely
at the level of the GTMDs and provide new information not contained in the
GPDs or TMDs. $F_{14}$ is related to the canonical OAM as shown in
\cite{lorce, hatta1, lorce2}:

\be
l^{q}_{z} = -\int dx d^{2}k_{\perp} \frac{k_{\perp}^2}{m^2} F_{14}.
\ee
 We give the final expression for the canonical quark OAM in the dressed quark model.
 \be \label{sl}
 l^{q}_{z} = -  2N \int dx (1-x^2)\Big[ I_{1} -
 m^2(x-1)^2 I_{2}\Big ]
 \ee%
The above expression is in agreement with \cite{hari}, where the authors have calculated the quark 
canonical OAM using the same model neglecting the quark mass. Our results
are  also in agreement with \cite{hikmat} as well as a recent calculation in
\cite{other}. We thus confirm the conclusion in \cite{other} in our model
calculation that the GTMDs $F_{14}$ and $G_{11}$ exist and non-zero, in contrast to the
arguments given in \cite{liuti}. Also in \cite{other} the above two GTMDs
were calculated incorporating the gauge link; as their results agree with
ours, it is clear that the gauge link does not contribute to these GTMDs and
the result is independent of the choice of the gauge link, which was also
noted in \cite{other}.     

As shown in \cite{lorce,lorce_14}, the correlation between the quark spin and its OAM
is given by

\be
C^{q}_{z} = \int dx d^{2}k_{\perp} \frac{k_{\perp}^2}{m^2} G_{11}.
\ee

As in our model $F_{14}$=$-G_{11}$, the above correlation is given by Eq.
(\ref{sl}). The spin-orbit correlation for the quark in the dressed quark is
negative. This is opposite to what is observed in chiral quark-soliton model
and constituent quark model, namely here the quark spin is anti-aligned with
its OAM, unlike the other two models where there is no gluon.

\begin{figure}[!htp]
\begin{minipage}[c]{1\textwidth}
\tiny{(a)}\includegraphics[width=8cm,height=7.5cm,clip]{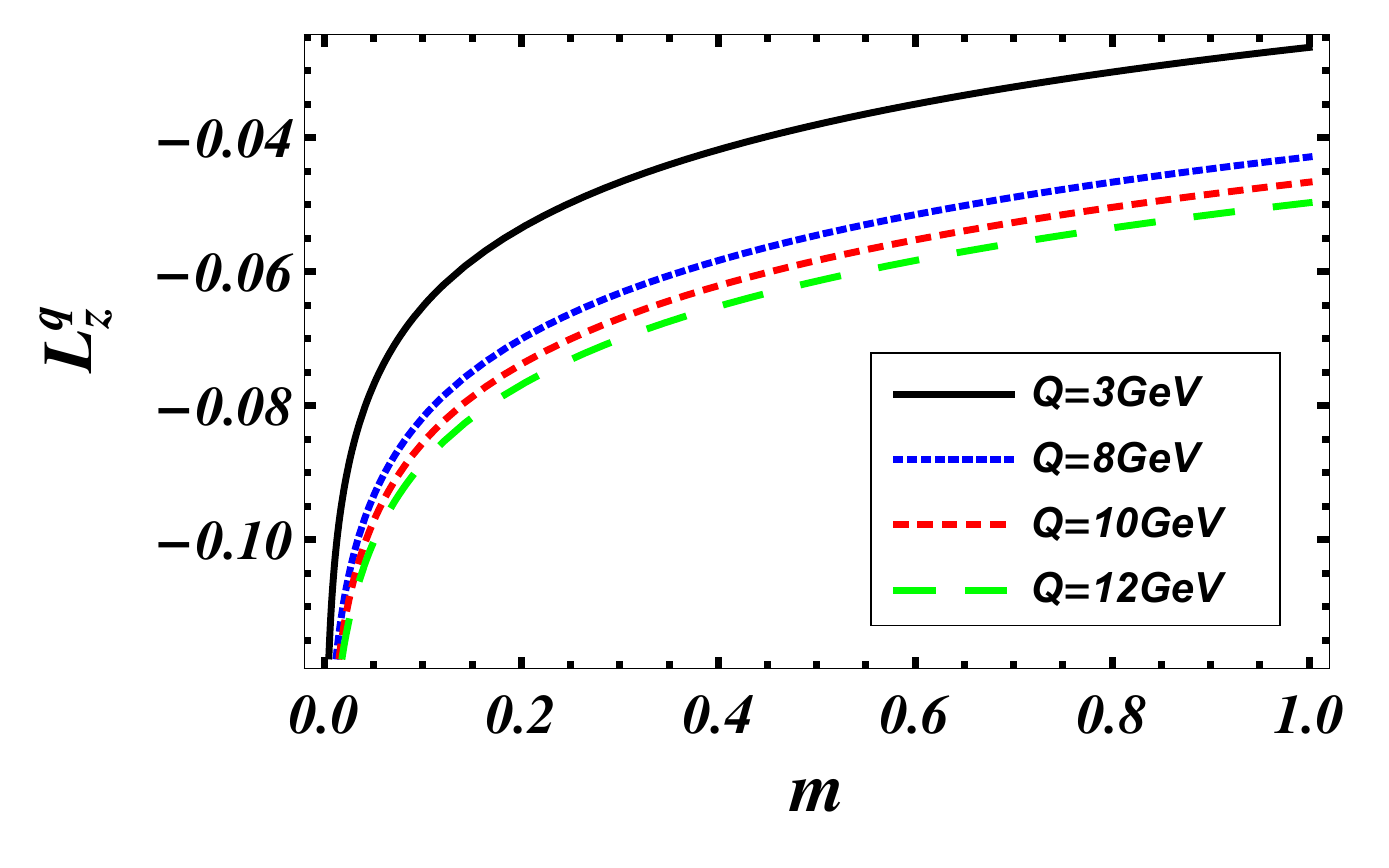}
\hspace{0.1cm}
\tiny{(b)}\includegraphics[width=8cm,height=7.5cm,clip]{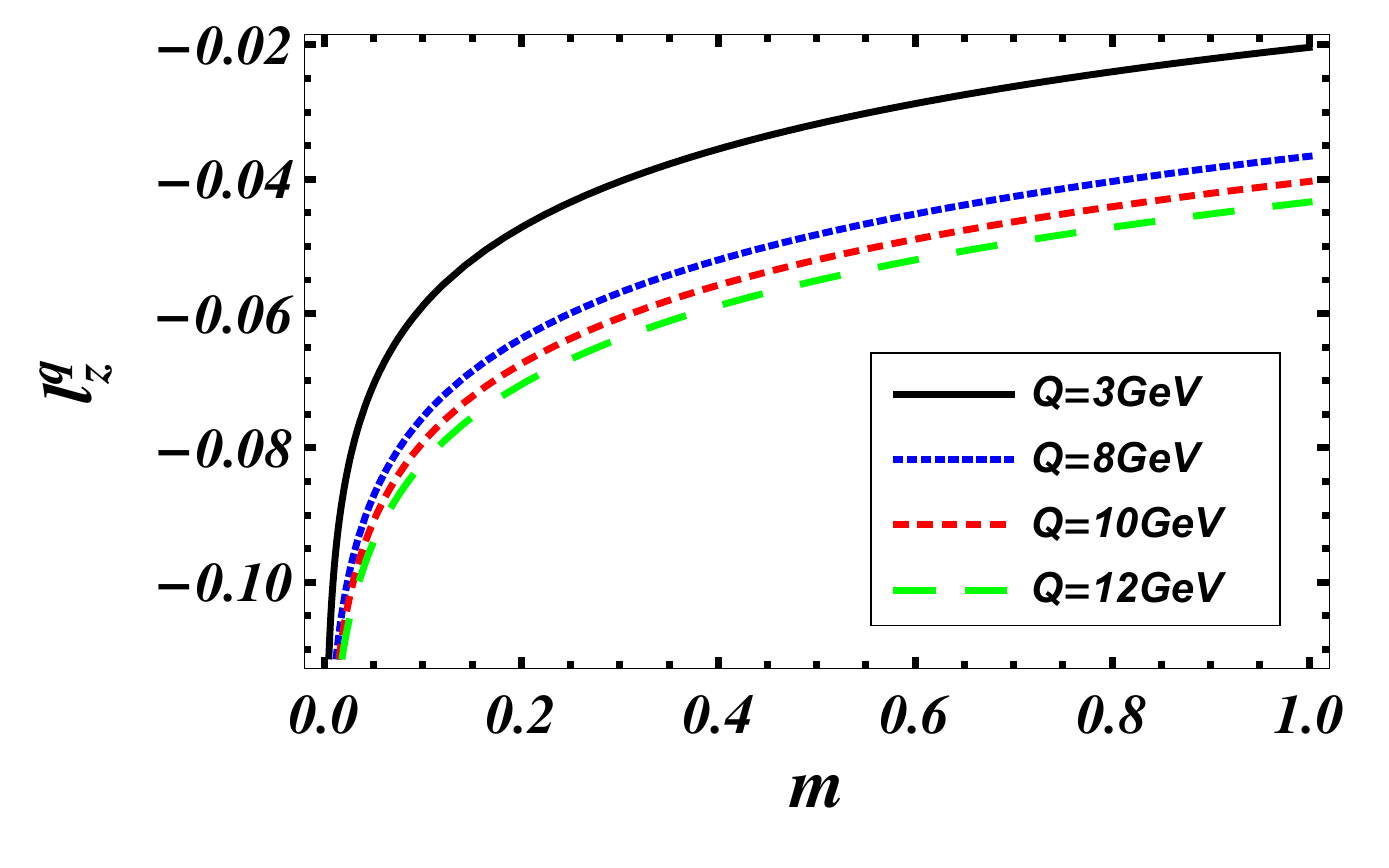}
\end{minipage}
\caption{\label{fig6}(Color online)
Plots of OAM (a) $L^{q}_z$ and (b) $l^{q}_z$ vs $m$  (GeV) for different values of $Q $
(GeV).
 }
\end{figure}

\section{Numerical Results}%

\noindent
In all plots, we have integrated over $x$ and divided by the normalization,
$N$. 
In Fig. \ref{fig1} we show the dependence of the Wigner distributions on the
quark mass. We took the mass of the dressed quark to be the same as the bare
quark. Here we have plotted the Wigner distributions versus the mass for fixed values
of $b_\perp$ in $GeV^{-1}$ and $k_\perp$ in $GeV$. Ideally the upper limit
of the $\Delta_\perp$ integration should be infinity. However we have
imposed an upper cutoff $\Delta_{max}$ in the numerical integration. In Fig.
1 we have taken $\Delta_{max} = 1.0$ GeV. Here $\vec{b_\perp} = b \hat{j}$ and 
$\vec{k_\perp} = k \hat{j}$. For $\rho_{UU}$ in Fig. 1 (a) we have plotted the mass dependence for three 
different values of $b_\perp$ which are 0.1, 0.5 and 1.0 $GeV^{-1}$ keeping  $k_\perp = 0.4$ GeV and we see that the value 
decreases with increasing mass. This is because the mass term in the denominator of Eq.(\ref{rhouu}) coming from the $D(k)$ 
function is dominant over the other term.
For larger  $b_\perp$ values the distribution has smaller values as seen from the plot. In Fig. 1 (b) we have plotted 
the mass dependence for three different values of $k_\perp$ which are 
0.1, 0.3 and 0.5 $GeV$ keeping  $b_\perp = 0.4$ $GeV^{-1}$.
 Again we see the same behavior as in Fig. 1 (a), in the lower mass range $\rho_{UU}$
increases sharply for smaller $k_\perp$. 
 In fig. 1 (c) and fig. 1 (d) we have plotted the mass dependence for  $\rho_{LU}$ with the same settings as for  $\rho_{UU}$. 
Since we choose  $\vec{k_\perp} = k \hat{j}$ and because of the factor
 $k_x \Delta_y - k_y \Delta_x $ we observe that the distribution has negative values but we do observe the same behavior 
as seen previously.
Lastly in Fig. 1 (e) and Fig. 1 (f)  we show the results for $\rho_{LL}$. Since $\rho_{UU}$  and $\rho_{LL}$ 
only differ by a sign in their mass term as seen in Eqs.(\ref{rhouu}-\ref{rholl}), 
the results are nearly identical, as the mass
term gives sub-dominant contribution.
In all the plots of \ref{fig1} we observe that at higher mass range the distributions are nearly independent 
of  $b_\perp$ and $k_\perp$ values.
 %
 %

\noindent 
In Fig. \ref{fig2} we show the 3D plots for the Wigner distribution $\rho_{UU}$. In 
the numerical calculation for Eq.\ref{rhouu} we have upper cut-off's $\Delta_x^{max}$ and 
$\Delta_y^{max}$ for the $\Delta_{\perp}$ integration.  In all plots we have taken $m=0.33$ GeV. 
In Figs. 2 (a) and (b) we have plotted $\rho_{UU}$ in $b$ space with $k_\perp = 0.4$ GeV such that  $\vec{k_\perp} 
= k \hat{j}$ for 
$\Delta_\perp^{max} = 1.0$ GeV and $\Delta_\perp^{max} = 5.0$ GeV respectively. 
We see that the plot has a peak centered at $b_x=b_y=0$ decreasing in the outer regions of the $b$ space.
 In \cite{lorce} the authors have shown that the contour plots show asymmetry associated with the 
orbital angular momentum and the asymmetry favored the $b \perp k$ direction to $b \parallel k$.
This can be understood from semi-classical arguments in a model with
confinement. As no confining potential is present in the perturbative model we consider here, the behavior 
is expected to be different. 
In our case we observe the asymmetry but there is no particular favored direction for this asymmetry.
In Figs. 2 (c) and (d) we have plots in the $k$ space where $b_\perp = 0.4$ GeV such that  $\vec{b_\perp} 
= b \hat{j}$ for $\Delta_\perp^{max} = 1.0$ GeV and $\Delta_\perp^{max} = 5.0$ GeV respectively. 
The behavior in the $k$ space is similar to that in the $b$ space but the peaks have negative values.
In Fig. 2 (e) and (f) we show the plots in the mixed space.  As discussed
earlier, Wigner distributions do not have probability interpretation due to
uncertainty principle in quantum mechanics. However in the distributions
$\rho_{UU} (k_y,b_x)$  we have integrated out the $k_x$ and $b_y$ dependence 
giving us the probability densities correlating $k_y$ and $b_x$, this
correlation is not restricted by uncertainty principle. Unlike in \cite{lorce} we observe a minima at 
$b_x=0$ and $k_y=0$. In fact the minima is observed for all $b_x$ values for $k_y=0$. 
As $\Delta_{max}$ increases the minima gets deeper. The plots show that the
probability of finding a quark with fixed $k_y$ and $b_x$ first increases
away from $k_y=0$ and then decreases.
  

\noindent 
\\ 
In Fig. \ref{fig3} we show the 3D plots for the Wigner distribution $\rho_{LU}$. 
This is the distortion of the Wigner distribution of unpolarized quarks due
to the longitudinal polarization of the dressed quark. In fig. 3 (a) and (b) we have 
plotted $\rho_{LU}$ in $b$ space with $k_\perp = 0.4 GeV$ such that  
$\vec{k_\perp} = k \hat{j}$ for $\Delta_\perp^{max} = 1.0$ GeV and $\Delta_\perp^{max} = 
5.0$ GeV respectively. Like in \cite{lorce} we observe a dipole structure in these plots 
and the dipole magnitude increases with increase in $\Delta_{max} $.
 In Fig. 3 (c) and (d) we have plots in the $k$ space where $b_\perp = 0.4$ GeV such that  $\vec{b_\perp} = 
b \hat{j}$ for $\Delta_\perp^{max} = 1.0 GeV$ and $\Delta_\perp^{max} = 5.0$ GeV respectively. 
 Again we observe a dipole structure but the orientation is rotated in the $k$ space when compared to the 
$b$ space plots of Fig. 3 (a) and Fig. 3 (b). As before 
 the dipole magnitude increases with increase in $\Delta_{max} $.
 In Fig. 3 (e) and (f) we show the plots in the mixed space. 
 We observe the quadrupole structure in the mixed space like in \cite{lorce} and the
 peaks increase in magnitude with increasing $\Delta_\perp^{max}$. 
 %
 %

\noindent 
\\ 
In Fig. \ref{fig4}  we show the 3D plots for the Wigner distribution $\rho_{LL}$. The behavior is similar to that of Fig \ref{fig2} since the
 Wigner distribution functions $\rho_{UU}$ and $\rho_{LL}$ only differ by the sign of the mass term in the numerator.\\
 %
 %

\noindent
In Fig.\ref{fig5} we have plotted the dependence of the Wigner distributions
on the upper limit of $\Delta_{\perp}$ integration. Ideally, the upper limit
of the FT should be infinite, but for practical purpose, a finite upper limit is
necessary. For physical processes, for example in the deeply virtual Compton scattering
(DVCS) such limits are there from the kinematics, that is the momentum
transfer should be much less than the virtuality of the photon, $Q$. Figs.
5(a), (b) and (c) show plots of $\rho_{UU}$, $\rho_{LU}$ and $\rho_{LL}$
respectively as functions of $ b_{\perp}$ for a fixed value of $k_\perp$
and different values of $\Delta_{\max}$. $\rho_{UU}$ and $\rho_{LL}$  show
similar behavior, which is expected from the analytic formulas. Both of them
show a peak at $\mid b_{\perp} \mid=0$, the peak becomes sharper as
$\Delta_{max}$ increases. $\rho_{LU}$ is zero at $b_\perp=0$ and changes
sign at the origin. Here we observe two peaks, and these move closer to $
\mid b_{\perp} \mid=0$ as $\Delta_{max}$ increases. This means that the
correlations between the unpolarized quarks inside the unpolarized target as
well as the distortions due to the longitudinal polarization of the quarks
in the longitudinally polarized dressed quark target are large in the close
vicinity of $b_\perp=0$ for fixed $k_\perp$. If the allowed transverse
momentum transfer is higher, these correlations move closer to the origin.
The distortions of the Wigner functions due to the longitudinal polarization
of the quark in an unpolarized target changes sign for negative $b_\perp$,
these distortions are related to the OAM of the quark. Such distortions are
also more concentrated near the origin in $b$ space as the transverse
momentum transfer is higher.  Similar conclusion can be drawn on the
spin-orbit correlation of the quark.
    
 %
 %

\noindent 
\\
In Fig.\ref{fig6} we have shown the orbital angular momentum of quarks as a function of the mass.
 Fig. 6 (a) is for $L^{q}_z$ and 6 (b) for $l^{q}_z$. Both the plots are shown for different
values of $Q$  
in GeV where $Q$ is the upper limit in the transverse momentum  integration.
As stated above, this is the large momentum scale involved in the process. 
 We see similar qualitative behavior of $L^{q}_z$ and $l^{q}_z$ where both are giving negative 
values for the  chosen domain of mass and also both the OAM decreases in
magnitude  with increasing mass.
However the magnitude of the two OAM differs in our model, unlike the case
in \cite{lorce}, where the same had been calculated in several models
without any gluonic degrees of freedom and  the total
quark contribution to the OAM were equal for both cases. It is to be noted
that there is only one quark flavor in the simple model we consider. In
\cite{lorce}, the contribution to the OAM from different quark flavors were
found to be different, but the sum over all flavors were equal for the two
definitions of OAM. Also, in \cite{hikmat} it has been shown that a
simple model without the gauge field (for example a scalar diquark model)
gives the same result for the above two definitions of quark OAM. Thus the
perturbative model we consider here explicitly shows the contribution of the 
gluonic degrees of freedom to the OAM, which has been calculated in  \cite{hari,other}.
In fact in \cite{hari} it was shown that in the model considered here,
after the inclusion of the single particle sector of the Fock space (which
contributes at $x=1$), the gluon intrinsic helicity contribution to the
helicity sum rule cancels the contribution from the canonical quark and gluon OAM and
the Jaffe-Manohar helicity sum rule is satisfied.
\section{Conclusion}
\noindent
In this work, we calculated the Wigner distributions for a quark state
dressed with a gluon using the overlap representation in terms of the LFWFs.
This is a simple composite spin-1/2 system which has a gluonic degree of
freedom. Although the Wigner distributions in quantum mechanics are not
measurable and do not have probabilistic interpretation, after integrating
out some of the variables a probabilistic interpretation is possible to
obtain. We calculated the Wigner distributions both for unpolarized and
longitudinally polarized target and quarks and showed the
correlations in transverse momentum and position space. We compared and contrasted the
results with an earlier calculation of Wigner distributions in light cone
constituent quark model and light-cone chiral quark soliton model. We also calculated 
the kinetic quark OAM using the
GPD sum rule and the canonical OAM and showed that these are different in
magnitude, the difference is an effect of the gluonic degree of freedom.       
We also found that in the limit of zero quark mass 
our result for the canonical OAM agrees with that of  \cite{hari}. We also
presented the results for the spin-orbit correlation of the quark.  
Further work would involve calculating the Wigner distributions for the
gluons and also including 
transverse polarization of the target and the quark.

\section{ACKNOWLEDGMENTS}
\noindent
We would like to thank C. Lorce and B. Pasquini for helpful discussion. This work is
supported by the  DST project SR/S2/HEP-029/2010, Govt. of
India. After we had put this paper in the arXiv, we became aware of
\cite{other} which appeared in the arXiv a couple of days before.

\end{document}